\documentclass{article}
\usepackage{subcaption}
\usepackage[english]{babel}
\usepackage{xcolor}

\usepackage[letterpaper,top=2cm,bottom=2cm,left=2.5cm,right=2.5cm,marginparwidth=1.75cm]{geometry}

\usepackage{amsmath}
\usepackage{dsfont}
\usepackage{amssymb}
\usepackage{graphicx}

\usepackage{stmaryrd}

\usepackage[colorlinks=true, allcolors=blue]{hyperref}
\usepackage[square, numbers, comma, sort&compress]{natbib}
\usepackage{authblk}
\usepackage{lineno}

\title{Mutant fate in spatially structured populations on graphs: connecting models to experiments}

\author{Alia Abbara\textsuperscript{1,2}, Lisa Pagani\textsuperscript{1,2,\footnote{Present address: Institute of Integrative Biology, ETH Zürich, Zürich, Switzerland}}, Celia García-Pareja\textsuperscript{1,2,\footnote{Present address: Department of Mathematics, School of Engineering Sciences, KTH Royal Institute of Technology, SE-10044 Stockholm, Sweden}}, Anne-Florence Bitbol\textsuperscript{1,2,\ddag}}
\affil{
\textbf{1} Institute of Bioengineering, School of Life Sciences, École Polytechnique Fédérale  de Lausanne (EPFL), CH-1015 Lausanne, Switzerland\\
\textbf{2} SIB Swiss Institute of Bioinformatics, CH-1015 Lausanne, Switzerland\\
\ddag Corresponding author: \href{mailto:anne-florence.bitbol@epfl.ch}{anne-florence.bitbol@epfl.ch}}
\date{}
\begin{document}
\maketitle

\begin{abstract}
In nature, most microbial populations have complex spatial structures that can affect their evolution. Evolutionary graph theory predicts that some spatial structures modelled by placing individuals on the nodes of a graph affect the probability that a mutant will fix. Evolution experiments are beginning to explicitly address the impact of graph structures on mutant fixation. However, the assumptions of evolutionary graph theory differ from the conditions of modern evolution experiments, making the comparison between theory and experiment challenging. Here, we aim to bridge this gap by using our new model of spatially structured populations. This model considers connected subpopulations that lie on the nodes of a graph, and allows asymmetric migrations. It can handle large populations, and explicitly models serial passage events with migrations, thus closely mimicking experimental conditions. We analyze recent experiments in light of this model. We suggest useful parameter regimes for future experiments, and we make quantitative predictions for these experiments. In particular, we propose experiments to directly test our recent prediction that the star graph with asymmetric migrations suppresses natural selection and can accelerate mutant fixation or extinction, compared to a well-mixed population. 
\end{abstract}

\section*{Author Summary}
Predicting how mutations spread through a population and eventually take over is important for understanding evolution. Complex spatial structures are ubiquitous in natural microbial populations, and can impact the fate of mutants. Theoretical models have been developed to describe this effect. They predict that some spatial structures have mutant fixation probabilities that differ from those of well-mixed populations. Experiments are beginning to probe these effects in the laboratory. However, there is a disconnect between models and experiments, because they consider different conditions. In this work, we connect them through a new model that closely matches experimental conditions. We analyze recent experiments and propose new ones that should allow testing the effects of complex population spatial structures on mutant fate.

\section*{Introduction}

Most natural microbial populations possess complex spatial structures. For example, pathogens evolve within each host and are transmitted between hosts during epidemics~\cite{Bertels19}, an important fraction of bacteria and archaea live in biofilms~\cite{Flemming19}, the gut microbiota occupies different habitats within the gut~\cite{Donaldson16} and evolves there~\cite{Garud19,Frazao22}, and the soil microbiota exhibits complex spatial patterns correlated with environmental factors~\cite{Mod21}. In this light, it is important to understand the impact of the spatial structure of populations on their evolution.

In population genetics theory, populations divided into multiple well-mixed demes (i.e.\ subpopulations) connected by migrations have long been studied~\cite{Wright31,Kimura64}. Most models assume that migrations are symmetric enough to maintain the overall mutant fraction. Under this hypothesis, the fixation probability of a mutant is not impacted by spatial structure~\cite{maruyama70, maruyama74}, unless deme extinctions occur~\cite{Barton93,whitlock2003}. Note however that several studies determining the effective population sizes of structured populations considered more general migration patterns~\cite{Nagylaki80,Whitlock97,Nordborg02,Sjodin05}, but their focus was mainly on neutral evolution~\cite{Barton93, Whitlock97,Nordborg02}. Evolutionary graph theory allows to model more complex spatial structures~\cite{lieberman2005evolutionary,Antal06}. It assumes that each individual is located on a node of a graph, and that birth and death events occur according to a specific update rule which ensures that exactly one individual sits on each node at all times. 
Evolutionary graph theory shows that some spatial structures can impact mutant fixation probability~\cite{lieberman2005evolutionary}. Different update rules yield different evolutionary outcomes~\cite{Hindersin15,tkadlec2020limits}. For instance, the star graph can amplify or suppress natural selection depending on the update rule~\cite{Kaveh15,Hindersin15,Pattni15}. More general models inspired by evolutionary graph theory assume that a well-mixed deme with constant size sits on each node of a graph, and use similar update rules as models with single individuals on nodes~\cite{Houchmandzadeh11,Houchmandzadeh13,Constable14,yagoobi2021fixation,Yagoobi23}. In these models, migration is either always coupled to birth and death events~\cite{Houchmandzadeh11,Houchmandzadeh13,Constable14,yagoobi2021fixation,Yagoobi23}, or independent from them but constrained to be symmetric~\cite{Yagoobi23}.

Modern evolution experiments~\cite{Lenski91,Elena03,Good17,Kryazhimskiy12,Nahum15,France19,Chen20,Chakraborty23,Kreger23} allow to investigate evolution in the lab. In particular, some experiments have investigated the impact of spatial structure on evolution, implementing spatial structure through transfers between well-mixed cultures~\cite{Kryazhimskiy12,Nahum15,France19,Chakraborty23,Kreger23}. A recent seminal study explicitly considered different graph structures, including the star graph~\cite{Chakraborty23}. In this exciting context, there is a strong disconnect between theoretical predictions and experiments. Experiments allow for complex structures and asymmetric migrations~\cite{Chakraborty23}, thus going beyond traditional population genetics models~\cite{Wright31,Kimura64,maruyama70,maruyama74}. Evolutionary graph theory assumes that the size of each deme is strictly constant (generally set to one, or to a fixed number in generalizations). This is not the case in experiments, which generally employ batch culture with serial passages (also called serial transfers), alternating phases of growth and bottlenecks~\cite{Lenski91,Elena03,Good17,Kryazhimskiy12,Nahum15,France19,Chen20,Chakraborty23,Kreger23}. Furthermore, these experiments do not have update rules or strict couplings between birth, death and migration events, in contrast to evolutionary graph theory. 

We recently developed a new model of spatially structured populations on graphs, which explicitly models serial passages with migrations, allows asymmetric migrations, and can handle large population sizes~\cite{Abbara23}. This makes it ideal to model current experiments, such as those performed in~\cite{Chakraborty23}, and to make quantitative predictions for future experiments. Formally, our model is close to a structured Wright-Fisher model, thus bridging classical population genetics models~\cite{Wright31,Kimura64,maruyama70, maruyama74,Barton93} and evolutionary graph theory~\cite{lieberman2005evolutionary,Hindersin15,tkadlec2020limits}. 

Here, we use our model to simulate and analyze the seminal experiments performed in~\cite{Chakraborty23} on the impact of complex spatial population structures on mutant fate. We find that in the conditions of this experiment, mutant fixation is systematic, and features intermediate plateaus corresponding to gradual mutant spread if migrations are relatively rare. We further show that the time evolution of the mutant fraction is well-described by a simple deterministic model. We conclude that no amplification or suppression of selection can be observed in this parameter regime. Next, we use our model to make quantitative suggestions for future experiments on the effect of spatial structure on mutant fate. We propose specific parameter regimes that should allow to observe suppression of selection and acceleration of fixation or extinction dynamics in the star compared to a well-mixed population. We make detailed and experimentally testable predictions in this parameter regime.

\section*{Results}

\subsection*{Modeling evolution experiments in spatially structured populations with serial passage}

\begin{figure}[htb!]
\begin{center}
\includegraphics[scale=0.6]{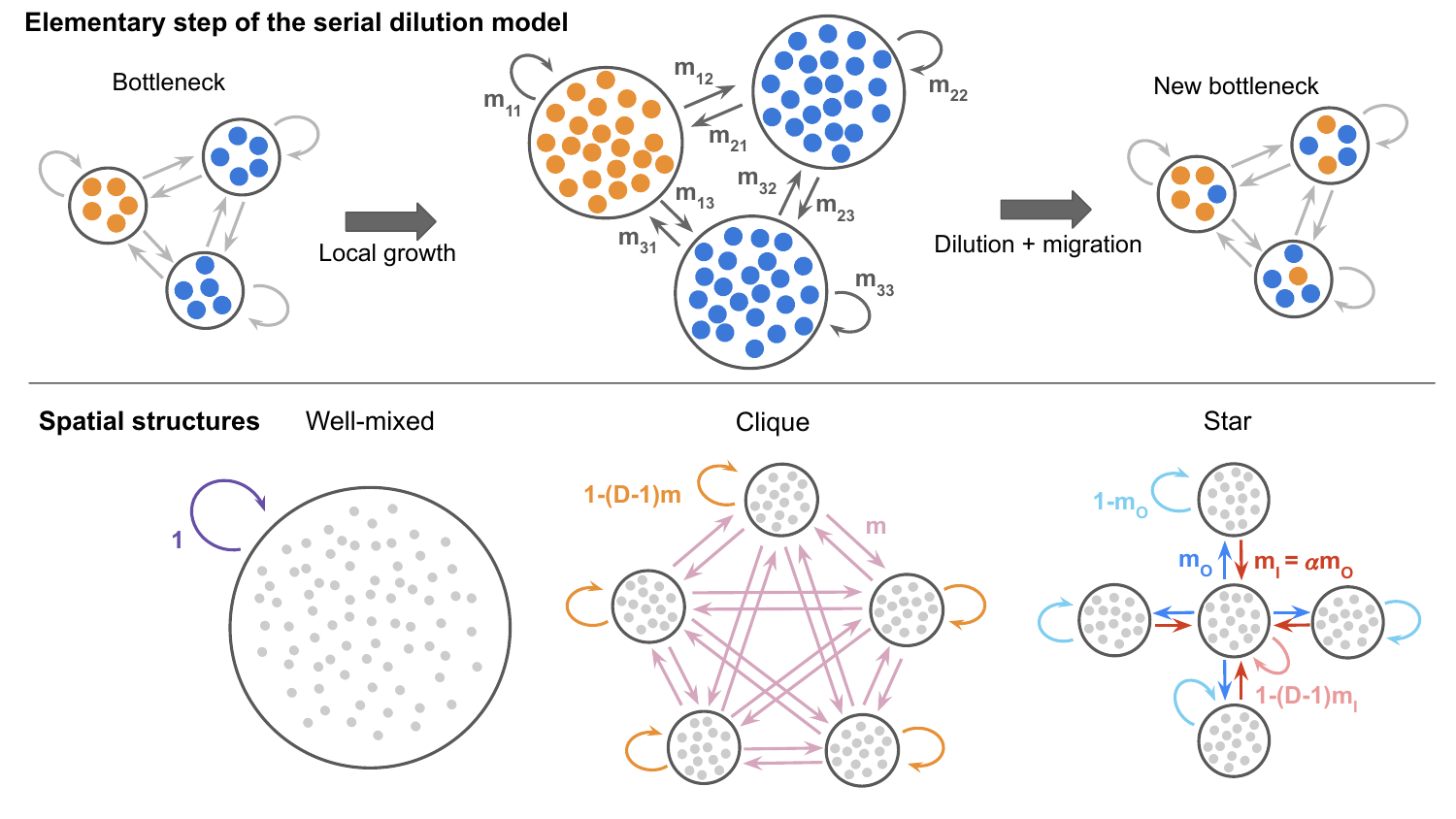}
\caption{\textbf{Model and spatial structures considered.} Top: schematic of one elementary step of our serial dilution model~\cite{Abbara23} for a structured population with three demes, initialized with one fully mutant (orange) deme. Starting from a bottleneck, each deme first undergoes a phase of growth. Then, dilution and migration, modeling serial passing with exchanges between demes, occur along the edges of the graph, according to migration probabilities. This yields a new bottleneck state.
Bottom: spatial structures considered. From left to right: a well-mixed population, the clique (or complete graph, or island model~\cite{Wright31}) and the star. Arrows represent migrations, and their probabilities are shown.
In the clique, migrations probabilities between different demes are all equal to $m$. In the star, the outgoing migration probability from the center to a leaf is $m_O$, and the incoming migration probability from a leaf to the center is $m_I = \alpha\, m_O$.}
\label{model_structures}
\end{center}
\end{figure}

\paragraph{Key ingredients and steps of the model.} In our recent paper~\cite{Abbara23}, we developed a model describing the evolution of spatially structured populations composed of well-mixed subpopulations or demes, each placed on a node of a graph. This model explicitly takes into account the serial passage protocol commonly used in modern evolution experiments~\cite{Lenski91,Elena03,Good17,Kryazhimskiy12,Nahum15,France19,Chen20,Chakraborty23,Kreger23}. In particular, such a protocol was used in the recent study~\cite{Chakraborty23} to investigate the impact of different spatial structures on the fate of beneficial mutants. In ``Model and methods", we describe our model step by step and we relate it to the experiments of~\cite{Chakraborty23}  (see also~\cite{Abbara23} for model details). Our model is illustrated in the top panel of Fig.~\ref{model_structures}, and the spatial structures we consider here are presented in the bottom panel of Fig.~\ref{model_structures}.

Briefly, we consider $D$ well-mixed demes, all with a bottleneck size that fluctuates around a value $B$, set by the serial passage protocol. In this spatial structure, two types of individuals are in competition: wild-types, and beneficial mutants that have a relative fitness advantage $s$ over wild-types.

Starting from a bottleneck, demes first undergo a phase of local exponential growth for a fixed time $t$. This time essentially represents the time between two successive serial passage events. As we do not take into account the possible slowdown of growth due to entry in stationary phase, $t$ in fact represents the time of exponential growth that would correspond to the actual growth between two successive serial passage events. It is during the growth phase that the fitness advantage of mutants matters: their growth rate is $(1+s)$ times larger than that of wild-types. As a result, in demes where mutants were present at the bottleneck, the fraction of mutants increases during the growth phase. This variation of mutant fraction only depends on the product $st$ of $s$ and $t$, which we will thus refer to as the \textit{effective fitness advantage} (see ``Model and methods"). 

Then, serial passage with spatial structure is performed through simultaneous dilution and migration events. Specifically, it is modeled as binomial samplings of wild-types and mutants from and to each deme, performed according to migration probabilities $m_{ij}$ set along each edge $(ij)$ of the graph representing the spatial structure. This serial passage step leads to a new bottleneck state in the spatially structured population. In this new bottleneck state, each deme again has a size around $B$, and is composed of individuals that originate from this deme and from others, according to the chosen migration probabilities. Note that our model incorporates fluctuations of bottleneck size and composition, consistently with experiments.

\paragraph{Link to other models, and summary of previous results.} Many traditional population genetics models assume that migrations are symmetric enough to preserve the overall mutant fraction in the population~\cite{Wright31,maruyama70,maruyama74}, a condition known as conservative migration~\cite{Blythe07}. In our framework, assuming that average bottleneck size $B$ is the same for each deme, i.e.\ $\sum_{j=1}^D m_{ji}=1$ for any deme $i$, and neglecting deme size variations, the conservative migration condition reads $\sum_i x'_i=\sum_{i,j}x'_i m_{ij}$ for all possible sets of mutant fractions $x'_i$ after growth. It leads to $\sum_j m_{ij}=1=\sum_j m_{ji}$ for all $i$. This corresponds to circulation graphs, where each deme has equal total incoming and outgoing migration flows. Note that the same conclusion holds assuming that $\sum_j m_{ij}=1$ for all $i$ instead of $\sum_j m_{ji}=1$ for all $i$. Our model allows us to consider more complex spatial structures, in the form of graphs that are not circulations. In other words, it allows us to consider asymmetric migrations that do not necessarily preserve overall mutant fraction. Furthermore, our model enables us to treat any connected graph with migration probabilities along its edges. It generalizes over evolutionary graph theory~\cite{lieberman2005evolutionary} and its extensions to nodes comprising multiple individuals~\cite{Houchmandzadeh11,Houchmandzadeh13,Constable14,yagoobi2021fixation,Yagoobi23}, which assume specific update rules that exactly preserve population size. 

We previously showed that our model recovers the predictions of evolutionary graph theory regarding amplification or suppression of selection in the star graph, in the limit of rare migrations and for specific migration asymmetries matching those implicitly imposed by update rules~\cite{marrec2021,Abbara23}. Furthermore, for circulation graphs, we showed that our model recovers Maruyama's theorem~\cite{maruyama70,maruyama74} of population genetics and generalizes the circulation theorem~\cite{lieberman2005evolutionary} of evolutionary graph theory, both for rare migrations~\cite{marrec2021} and for frequent migrations in the branching process regime~\cite{Abbara23}.

\subsection*{Analysis of the experiments performed in~\cite{Chakraborty23}}

Recent innovative experiments described in~\cite{Chakraborty23} tracked the fraction of an antibiotic resistant mutant with substantial fitness advantage, as it spreads through a star or a clique network, with various migration intensities. Our serial dilution model is close to the experimental protocol from~\cite{Chakraborty23}, as both rely on successive steps of local growth in each node followed by bottlenecks with exchanges between demes. In~\cite{Chakraborty23}, it was found that the overall mutant fraction, measured at the bottleneck, can grow faster in a star structure than in a clique, when mutants are initially placed in a leaf, under relatively rare migrations. Measurements of the mutant fraction were performed during 7 to 9 bottlenecks, and show important variability across replicates.

What does our model predict for the experiments performed in~\cite{Chakraborty23}, in particular for their long-term outcome? To address this question, we perform numerical simulations of our model with parameters corresponding to the experiments of~\cite{Chakraborty23}, see ``Model and methods" for details of these parameter choices. In particular, there are $D=4$ demes, each with average bottleneck size $B=10^7$, one of them is initialized with $10^4$ mutants out of $10^7$ individuals, the selective advantage of mutants is $s=0.2$ and the growth time is $t=\log(100)$. Importantly, with such a large effective selective advantage and such a large initial number of mutants, mutant fixation would be certain both in an isolated deme and in a well-mixed population with $DB$ individuals, as shown by the fixation probability in Eq.~\ref{fix} of ``Model and methods".

Our serial dilution model usually assumes soft selection (see ``Model and methods" for details). Indeed, the total size of demes after growth does not directly affect their contributions to the next bottleneck state~\cite{Wallace75}. 
It is challenging to exactly map the experiments in~\cite{Chakraborty23} to soft or hard selection. Having complete growth curves for both the mutant and the wild-type in the experimental conditions would be very valuable to this end. However, an experimentally observed difference in optical density of the mutant and non-mutant demes after growth~\cite{ChakrabortyP} rather points to hard selection, given the fixed dilution factor employed~\cite{Chakraborty23}. Thus, we extended our model to implement hard selection. All our main conclusions are robust to considering either version of the model. Therefore, we focus in the main text on the soft selection version of our model, as in~\cite{Abbara23}. The hard selection version is presented in Section~\ref{Section_hard_selection} and Fig.~\ref{hard_selection_fig} of the Supplement.

\paragraph{Choosing the reference: clique or well-mixed population.}
Previous theoretical works have compared the fixation of a mutant in a structured population to that in \emph{well-mixed} populations, with the same total number of individuals, but placed in a unique deme, and with the same initial overall mutant fraction~\cite{maruyama70,maruyama74,marrec2021,Abbara23}. Indeed, taking the well-mixed population as reference allows to best assess the impact of spatial structure. In our previous work~\cite{Abbara23}, we further considered a graph structure called the \emph{clique}, also known as the island model~\cite{Wright31}, which is a fully connected graph with equal migration intensities between all demes (see Fig.~\ref{model_structures}). Due to its strong symmetry, this structure displays some remarkable properties. In particular, the fixation probability of a mutant in a clique is the same as in a well-mixed population~\cite{maruyama70,maruyama74,Abbara23}. In~\cite{Chakraborty23}, the term ``well-mixed'' was used to refer to a clique, and the evolution of mutant fraction was tracked in two structured populations: the star and the clique. Choosing the clique rather than a true well-mixed population as a reference is experimentally relevant because bacterial physiology and growth are impacted by the shape, size, and volume of the culture flask~\cite{Munch2020, Pedersen2021, Kram2014}. Therefore, several factors other than spatial structure would need to be taken into account to analyze experiments comparing mutant growth in a homogeneous well-mixed population to that in several flasks with smaller volumes. In the following, we retain the terminology of~\cite{Abbara23}, referring to the fully connected graph as a clique. For the sake of completeness, we show the results of our model for both reference structures, namely the well-mixed population and the clique, even though only the clique was considered in experiments~\cite{Chakraborty23}.

\begin{figure}[htb!]
    \centering
\includegraphics[scale=0.42]{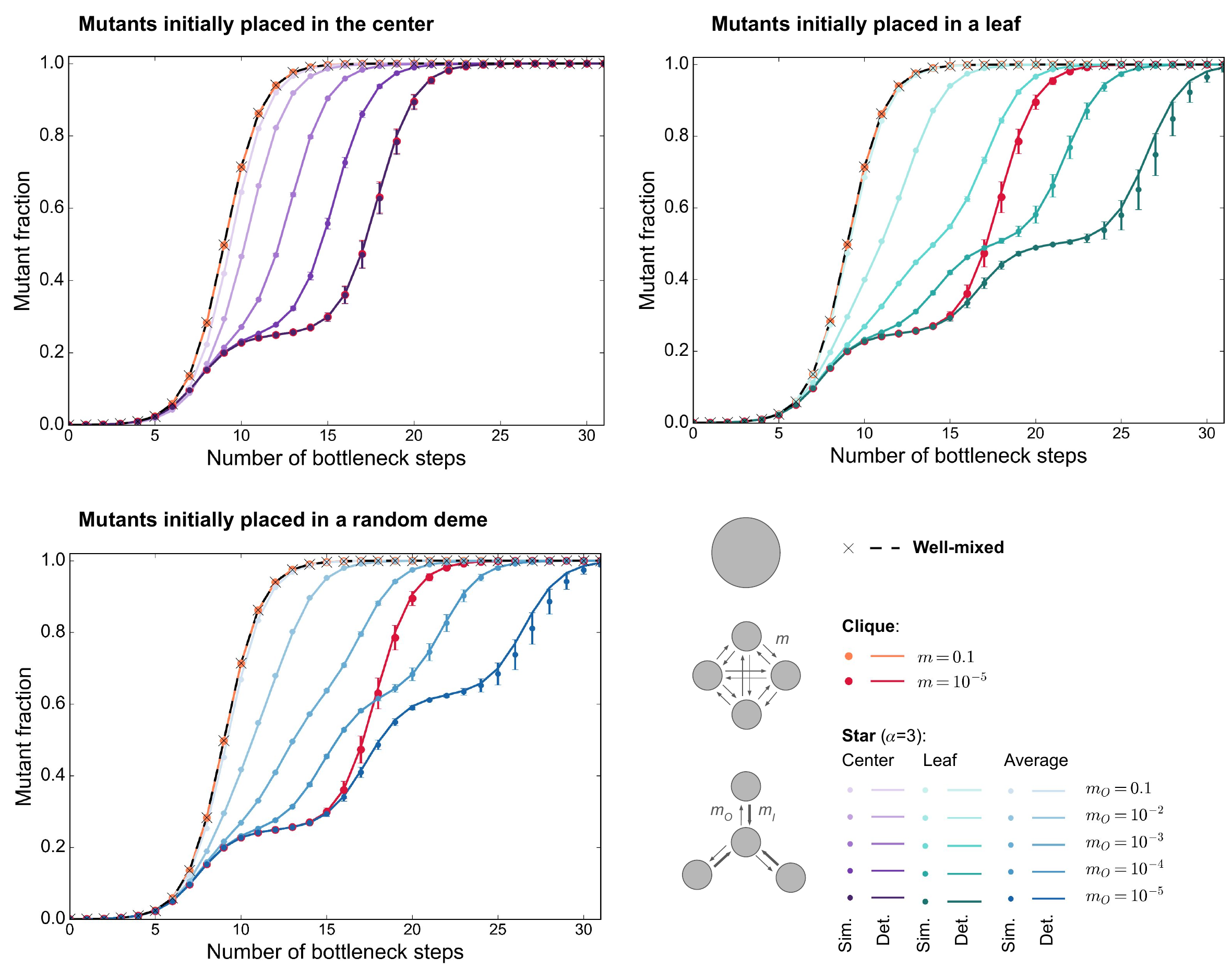}
    \caption{\textbf{Growth of the mutant fraction in different structures.} As in~\cite{Chakraborty23}, we consider a clique and a star with $D=4$ demes of bottleneck size $B=10^7$ each, where $10^4$ mutants with effective fitness advantage $st=0.2 \log(100)$ are initially placed in one deme. This deme is the center (top left), a leaf (top right), or chosen uniformly at random (bottom). As a reference, we also consider a well-mixed population of total bottleneck size $DB=4 \times 10^7$, initialized with $10^4$ mutants. In each case, we report the mutant fraction versus time (expressed in number of bottleneck steps). We consider different migration probabilities $m$ for the clique and $m_O$ for the star, keeping $\alpha=m_I/m_O=3$ for the star. Markers are simulation results averaged over 100 trajectories, and error bars report the standard deviation. Note that all trajectories resulted in mutant fixation. Lines show the predictions from our deterministic model.}
    \label{fixation_det}
\end{figure}

\paragraph{Systematic mutant fixation with intermediate plateaus.} Fig.~\ref{fixation_det} shows our simulation results for the time evolution of the mutant fraction in the clique, and in the star with migration asymmetry $\alpha=m_I/m_O$ set to $3$ (see Fig.~\ref{model_structures}, bottom panel). 
Our top right panel corresponds to the experimental results shown in  Fig.~3 of~\cite{Chakraborty23}, in the ``IN$>$OUT'' setting. The migration probability $m_O$ in our Fig.~\ref{model_structures} corresponds to their migration intensity: for instance, $m_O=10^{-5}$ means that their migration intensity is equal to 0.001\%.

We observe that in all cases, fixation is reached. This is due to the large effective selective advantage and the large initial number of mutants.

Furthermore, Fig.~\ref{fixation_det} shows that, if migrations are frequent, the time evolution of mutant fraction in a structured population is extremely similar to that in a well-mixed population with size $DB$. This holds both in the clique and in the star. For less frequent migrations, fixation becomes slower, because it takes time for mutants to migrate from their initial demes to other ones. For the rarest migrations considered, we observe intermediate plateaus of mutant fraction, which correspond to successive fixation events in each deme. This is qualitatively reminiscent of the rare migration regime~\cite{marrec2021,Abbara23}, where fixation occurs sequentially. Note however that our theoretical definition of the rare migration regime is more stringent, as it requires that any migration (even of a single cell) is rare during fixation~\cite{marrec2021,Abbara23}. Both for the clique and for the star, a first plateau with mutant fraction 0.25 is reached, corresponding to one deme out of four having fixed the mutant, while others are still wild-type. We checked that this deme is the one where mutants started. This is consistent with the experimental observations in~\cite{Chakraborty23}, where plateaus are observed in the low migration intensity regime in Fig.\ 3. After this first step, trajectories differ for the clique and for the star, if the mutants starts from a leaf of the star (which is the case in~\cite{Chakraborty23}). Indeed, in the clique, mutants can migrate to any other deme from the one where mutants started. Meanwhile, in the star, if mutants started in a leaf, they first spread to the center before being able to spread to other leaves, which leads to a second plateau. Thus, in the star, we observe no second plateau if mutants start in the center, while a second plateau exists at mutant fraction 0.5 if mutants start in a leaf, see Fig.~\ref{fixation_det}. Note that, when mutants start in the center, the overall trajectory looks identical to that observed in the clique. Indeed, once the starting deme is dominated by mutants, spreading to all leaves in the star is as fast as spreading to all other demes in the clique, since we chose $m=m_O$, following Ref.~\cite{Chakraborty23}. When choosing uniformly at random the deme where mutants start, we thus obtain a plateau around mutant fraction $0.5\times 3/4+1/4=0.625$. Note that in~\cite{Chakraborty23}, only the first plateau was observed because experiments were then stopped.

\paragraph{Quasi-deterministic behavior.} An important observation in Fig.~\ref{fixation_det} is that there is little variability between trajectories. Indeed, standard deviations are small. Accordingly, examining individual trajectories confirms that they are rather similar, see Fig.~\ref{fixation_det_leaf_center}. Thus motivated, we proposed a simple deterministic model for the time evolution of mutant fraction (see ``Model and methods"). Fig.~\ref{fixation_det} shows a good overall agreement between the predictions of this deterministic model and the average of stochastic simulations. Thus, in the regime considered in~\cite{Chakraborty23}, the main features of the growth of mutant fraction are well-captured by a simple deterministic model. Figs.~\ref{fixation_det} and~\ref{fixation_det_leaf_center} also show that the variability across trajectories becomes larger when migrations are rarer, as the stochasticity of individual migration events starts to matter. The largest variability is observed at the last step of mutant spread starting from a leaf in the star (Fig.~\ref{fixation_det_leaf_center}, right panel), when the randomness of relatively rare migrations compounds over two steps of deme invasion. A slight discrepancy between deterministic predictions and the average of stochastic simulations is visible at this last step, when stochasticity becomes more important (see Fig.~\ref{fixation_det}).

Since fixation is certain and the time evolution of mutant fraction is essentially deterministic in the parameter regime considered in~\cite{Chakraborty23}, we do not predict amplification or suppression of natural selection for these parameters. Whatever the spatial structure, fixation probability is extremely close to one. Furthermore, Fig.~\ref{fixation_det} shows that fixation is slower in the clique and in the star than in the well-mixed population. Here, spatial structure slows down mutant fixation, all the more that migrations are rare. We do not find that the mutant fraction increases faster in the star starting from a leaf than in the clique with similar migration probabilities. This is at variance with the observations of~\cite{Chakraborty23} for the rarest migrations considered, suggesting that other effects were at play in these experiments. Note however that a faster increase of mutant fraction in the star than in the clique can be observed from the first plateau for mutants starting from the center if $m_O>m$, where $m$ is the migration probability between two demes in the clique (see Fig.~\ref{alpha1/3}). Indeed, this allows them to spread faster from their initial deme than in the clique. This would also be the case for mutants starting from a leaf if $m_I>3m$. 

The results in Fig.~\ref{fixation_det} correspond to $\alpha=3$. Ref.~\cite{Chakraborty23} explored the impact of various migration asymmetries $\alpha$ and also considered smaller bottleneck deme sizes $B=10^5$. For completeness, we show our results for $B=10^7$ and $\alpha=1/3$ in Fig.~\ref{alpha1/3}, and for $B=10^5$ and $\alpha=3$ in Fig.~\ref{alpha3_small}. 
 These two figures correspond respectively to Fig.~3 from~\cite{Chakraborty23}, in the ``OUT$>$IN'' setting, for high (1\%) and low (0.001\%) migration intensities, and to Fig. 6 from~\cite{Chakraborty23}. In these cases, our results are qualitatively similar to those of Fig.~\ref{fixation_det}, but stochasticity becomes more important for the smaller population size $B=10^5$.


\subsection*{Suggestions for experiments on the effect of spatial structure on mutant fate}

\paragraph{Goals and relevant regimes.} How to best choose experimental conditions to investigate the impact of spatial structure on mutant fate? Our model allows us to directly address this question. It is of particular interest to design experiments that may evidence amplification or suppression of natural selection by spatial structure. Given the experimental challenges associated with comparing the well-mixed population to structured ones (see above), these effects are defined by comparing the probability of mutation fixation in a spatially structured population to those observed in the  clique. Amplification (resp.\ suppression) means that beneficial mutations are more likely (resp.\ less likely) to fix in the structured population than in the  clique, and vice-versa for deleterious mutations. 

To observe these effects, it is best to choose parameter regimes where fixation probabilities in the well-mixed population and in  the clique  are intermediate. With beneficial mutants, this requires starting with few mutants, so that extinction can occur due to stochastic fluctuations. Alternatively, one can consider small effective selective advantages, even with larger initial numbers of mutants -- neutral mutants have a fixation probability equal to their initial fraction. These conditions can be checked in the fixation probability in Eq.~\ref{fix} of ``Model and methods": if the product of the initial number $n$ of mutants and of the effective mutant fitness advantage $st$ is large, the fixation probability becomes close to one. Thus, to have an intermediate fixation probability, this product needs to be small. Starting with small but controlled initial numbers of mutants (ideally, one single mutant) is experimentally challenging and would result in mutant extinction in a large fraction of experimental replicates. For simplicity, we propose experiments starting from one fully mutant deme, but with small mutant fitness advantages. However, small fitness differences entail that trajectories will feature strong fluctuations, and that fixation or extinction will take a long time (this would also be a concern starting with few mutants). For experiments not to be overly long, we suggest using small bottleneck sizes. Small populations also have the advantage that \textit{de novo} mutations are rarer than in large populations, allowing to observe the dynamics of preexisting mutant lineages without interference for a longer time.

\begin{figure}[htb!]
    \centering
\includegraphics[scale=0.41]{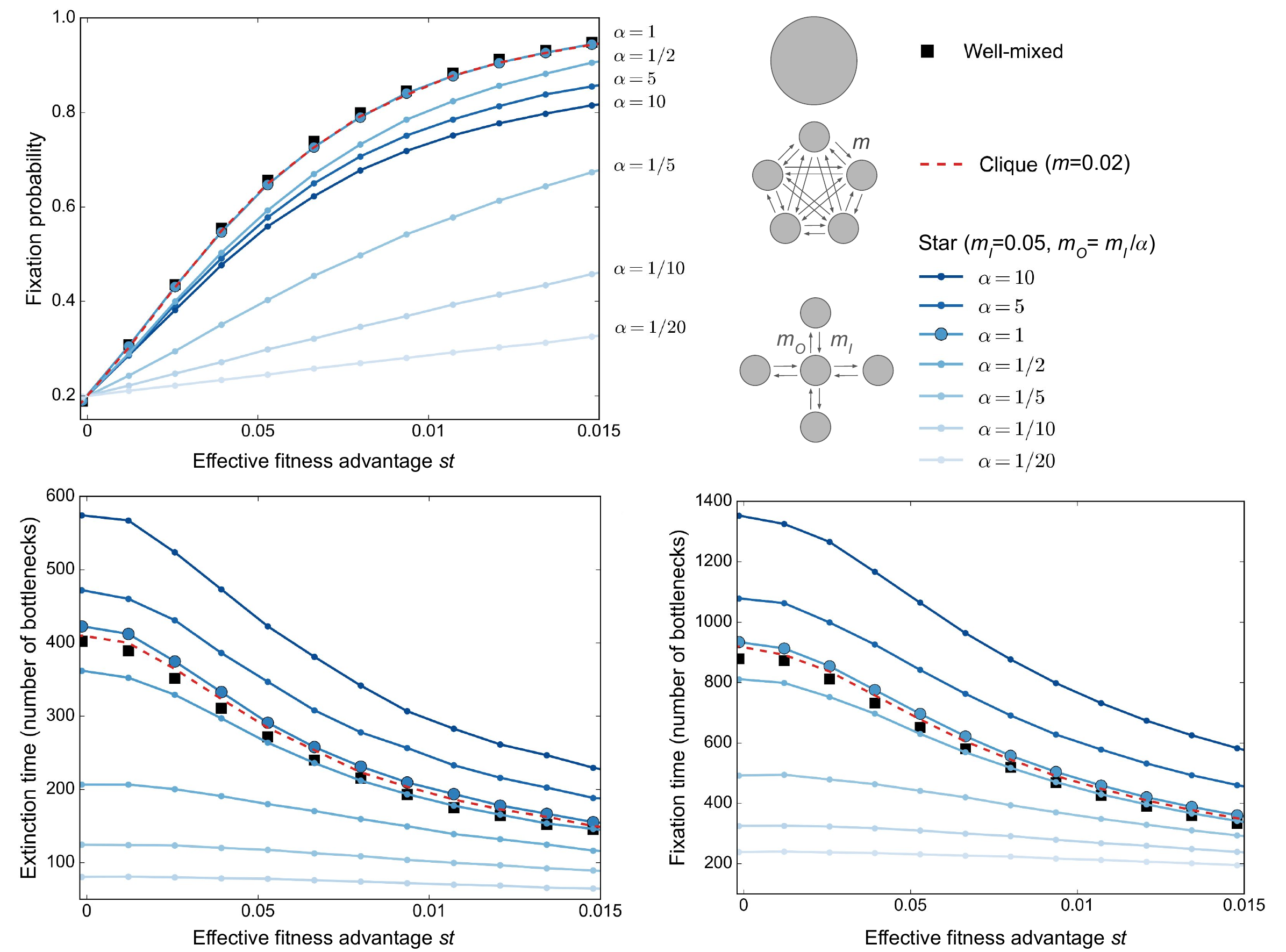}

     \caption{\textbf{Mutant fixation and extinction in different structures.} Mutant fixation probability (top), average extinction time (bottom left) and average fixation time (bottom right) are plotted as a function of the effective fitness advantage $st$ of the mutant. We consider a star and a clique with $D=5$ demes of size $B=100$, initialized with one fully mutant deme chosen uniformly at random. For reference, we also consider a well-mixed population of size $DB=500$, initialized with 100 mutants. For the star, we take different values of $\alpha=m_I/m_O$, always with $m_I=0.05$. For the clique, we choose $m=0.02$ so that the sum of all exchanges between demes is the same in the clique as in the star for $\alpha=1$. Markers are simulation results, obtained over at least 100,000 realizations. Lines linking markers are guides for the eye.  The values of $\alpha$ are indicated to the right of the top panel, highlighting the non-monotonous order of the plots.}
    \label{fprobas_etimes_ftimes_full}
\end{figure}

\paragraph{The star with asymmetric migrations suppresses selection and can accelerate fixation and extinction.} Let us thus investigate fixation probabilities and fixation times in the star in this regime of parameters. We choose $B=100$ as the average deme bottleneck size, and we take $D=5$ demes. Note that increasing $D$ increases the difference between the star and the clique, but large $D$ would make experiments more technically involved, and $D=4$ was considered in~\cite{Chakraborty23}. Note also that bottleneck sizes are often much larger in evolution experiments, but populations with bottleneck sizes of 5000 were considered in~\cite{Kryazhimskiy12}. We also choose migration probabilities that are not too small to avoid slowing down the dynamics. In practice, we take $m_I=0.05$. Fig.~\ref{fprobas_etimes_ftimes_full}, top panel, shows the fixation probability versus the effective fitness advantage $st$ in the star, starting from one fully mutant deme, for different migration asymmetries $\alpha=m_I/m_O$. We compare with results for the clique and the well-mixed population. We find that beneficial mutants are less likely to fix in the star than in the well-mixed population or in the clique: the star suppresses natural selection for $\alpha\neq 1$. This is consistent with our previous results in the branching process regime, where mutant fractions are small, while $B\gg 1$, $1/B\ll st\ll 1$, and migrations are not too rare~\cite{Abbara23}. Here, we observe that they extend beyond this regime. Note that for $\alpha=1$, the star behaves very similarly to a well-mixed population. The star is then a circulation~\cite{lieberman2005evolutionary}, and this is consistent with our results for such graphs~\cite{Abbara23}. Note also that the predictions of evolutionary graph theory, including the amplifier behavior of the star for some migration asymmetries matching the Birth-death update rules, are recovered in the rare migration regime~\cite{marrec2021,Abbara23}. However, rare migrations are problematic for experiments because mutant spread is then very slow, so we do not consider them here. We further observe that suppression of selection is stronger if $\alpha$ is smaller. 

Fig.~\ref{fprobas_etimes_ftimes_full}, bottom panels, shows the extinction time and fixation time versus the effective fitness advantage $st$. We compare the extinction and fixation times in the star to those in the well-mixed population and in the clique with the same total migration intensity between demes. We observe that both extinction and fixation are slower in the star for large values of $\alpha$, and faster for small values of $\alpha$. Because here we varied $\alpha$ at constant $m_I$ by setting $m_O=m_I/\alpha$, large $\alpha$ means small $m_O$, which slows down mutant spread. The acceleration observed for smaller $\alpha$ and larger migration probabilities is consistent with our previous results in the branching process regime~\cite{Abbara23}. 
However, this effect was not predicted by evolutionary graph theory, where each node of the graph structure is occupied by a single individual. Indeed, the fixation time for a mutant is slower in other graphs than in the clique of the same size~\cite{hindersin2013fixation,hindersin2014counterintuitive,Askari2015}, and scales differently with graph size~\cite{sharma2023self}.

Based on Fig.~\ref{fprobas_etimes_ftimes_full}, the best choice to observe a strong effect of spatial structure on mutant fate is to take small values of $\alpha$. This also has the advantage of minimizing experimental duration, which nevertheless remains large -- on the order of a few hundreds of bottleneck steps to see fixation or extinction. 

The results shown in Fig.~\ref{fprobas_etimes_ftimes_full} are based on considering an initial mutant deme chosen uniformly at random. We also report results obtained when starting from a mutant center, see Fig.~\ref{fprobas_etimes_ftimes_fullcenter}, or a mutant leaf, see Fig.~\ref{fprobas_etimes_ftimes_fullside}. In particular, for small $\alpha$, the fixation probability starting from a mutant center is larger than in a well-mixed population or in the clique, while the opposite holds when starting from a mutant leaf. Experiments contrasting these two cases would be interesting.

We emphasize that our predictions of suppression of selection and of acceleration of extinction and of fixation are really associated to the spatial structure of the star, and more precisely, to the asymmetry of migrations represented by $\alpha$. Indeed, in the clique, no such effect is observed, see Fig.~\ref{etimes_ftimes_full_clique}. Note that the clique is a circulation graph (as well as the star with $\alpha=1$), and has a fixation probability which is very close to that of the well-mixed population. In fact, in the branching process regime, we showed that all circulations have the same fixation probability as the well-mixed population~\cite{Abbara23}. However, simulations show that fixation is slower the clique than in the well-mixed population, as long as migration intensity is small. When migration intensity becomes large enough, our simulation results suggest that circulations have the same fixation time as the well-mixed population.

\paragraph{Detailed predictions regarding mutant fixation.} Let us predict what could be observed in experiments in this regime of parameters. For this, we consider the star with the same parameters as above ($B=100$, $D=5$, $m_I=0.05$ and $\alpha=1/10$, entailing $m_O=0.5$). We set the effective fitness advantage of the mutants to $st=0.01$, which exhibits both strong suppression and strong acceleration, see Fig.~\ref{fprobas_etimes_ftimes_full}. Fixation probabilities are the following in the star: $\rho=0.96$ starting from a mutant center, $\rho=0.23$ starting from a mutant leaf, and $\rho=0.38$ starting from a uniformly chosen mutant deme. For comparison, the probability of fixation is $\rho=0.86$ in the clique, as well as in the well-mixed population. These fixation probabilities reveal a substantial suppression of selection in the star (starting from a uniformly chosen mutant deme). Recall that these values were estimated from 100,000 replicate numerical simulations (see Fig.~\ref{fprobas_etimes_ftimes_full}). Importantly, the suppression effect should be detectable by performing at least 16 experimental replicates for each spatial structure. Indeed, the 95\% confidence intervals for the fixation probabilities in the star and in the well-mixed population, estimated using the Wilson binomial proportion confidence interval~\cite{Wilson27,Brown01}, should then become disjoint (see Methods for details).

\begin{figure}[htbp]
        \centering
\includegraphics[scale=0.48]{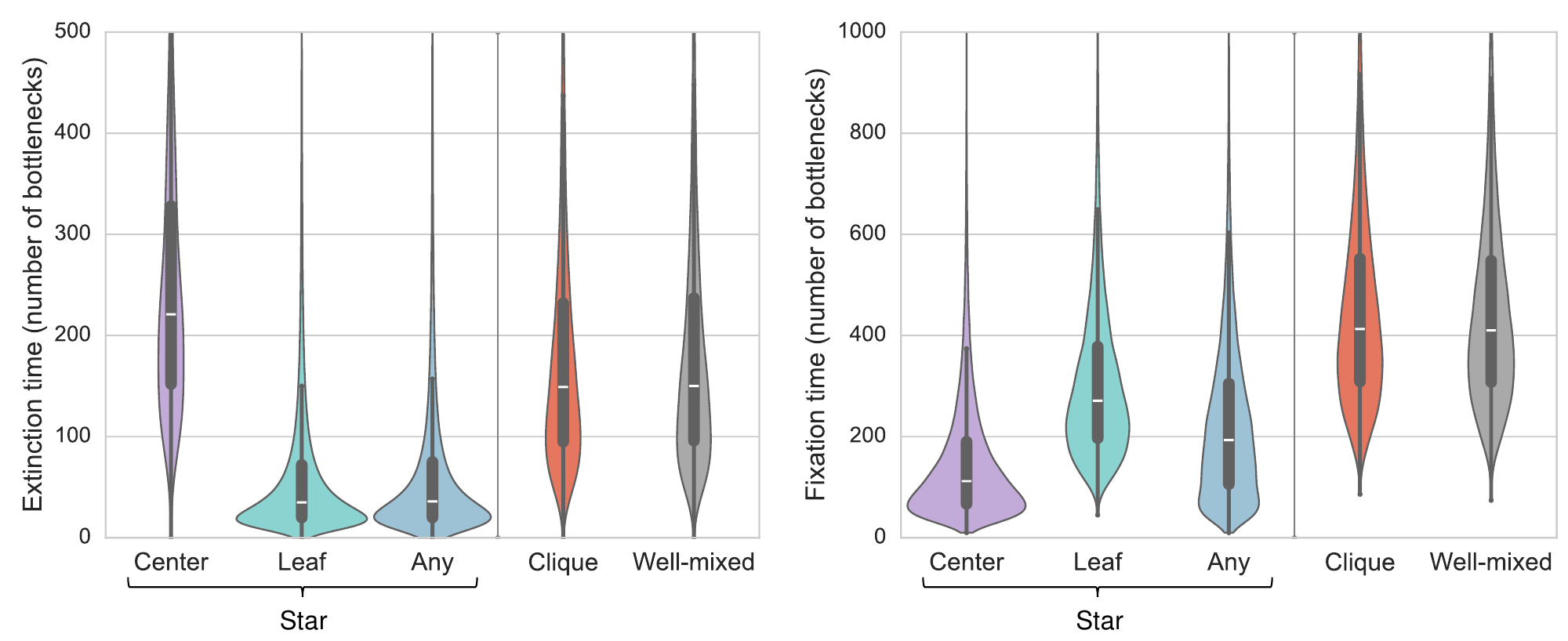}

    \caption{\textbf{Distributions of extinction and fixation times in different structures.} Violin plots report the distributions of extinction times in extinction trajectories (left), and of fixation times in fixation trajectories (right). Inside each violin plot, box plots report the median (white marker) and interquartile ranges (thick black bar). Times are expressed in number of bottleneck steps. Each violin plot is derived from at least 5000 trajectories. The star and clique comprise $D=5$ demes of size $B=100$ each, and  are initialized with one fully mutant deme. For the star, this deme is either the center, or a leaf, or a uniformly chosen one (``Any"). The well-mixed population comprises $DB=500$ individuals, and is initialized with 100 mutants and 400 wild-types. The effective fitness advantage of the mutants is set to $st=0.01$. For the star, $m_I=0.05$ as in Fig.~\ref{fprobas_etimes_ftimes_full}, and $\alpha=1/10$. For the clique, we choose $m=0.11$ so that the sum of all exchanges between demes is the same in the clique as in the star for $\alpha=1/10$ considered here. }
    \label{violin}
    \end{figure}

Fixation and extinction times can be measured in the same experiments as fixation probabilities, by counting how many serial passage steps are needed to get to fixation or extinction in each trajectory. An important point is that trajectories exhibit substantial variability in this regime of parameters. Fig.~\ref{violin} reports the distributions of fixation and extinction times in the star, in the clique and in the well-mixed population. Starting from the leaf or from a uniformly chosen mutant deme, the star accelerates the dynamics of extinction and of fixation. Conversely, starting from a mutant center makes extinction slower, because of the large outgoing migration probability $m_O$. All extinction and fixation times are of the order of some hundreds of bottleneck steps at most. The acceleration of fixation starting from a mutant center in the star, compared to the clique, should be detectable by performing 9 experimental replicates in each case. Indeed, the 95\% confidence intervals on the mean fixation times become disjoint for about 7 trajectories leading to fixation in each case (see Methods for details), and about 9 trajectories should be needed to have 7 of them leading to fixation in these cases.

\begin{figure}[htb!]        
        \centering
\includegraphics[scale=0.68]{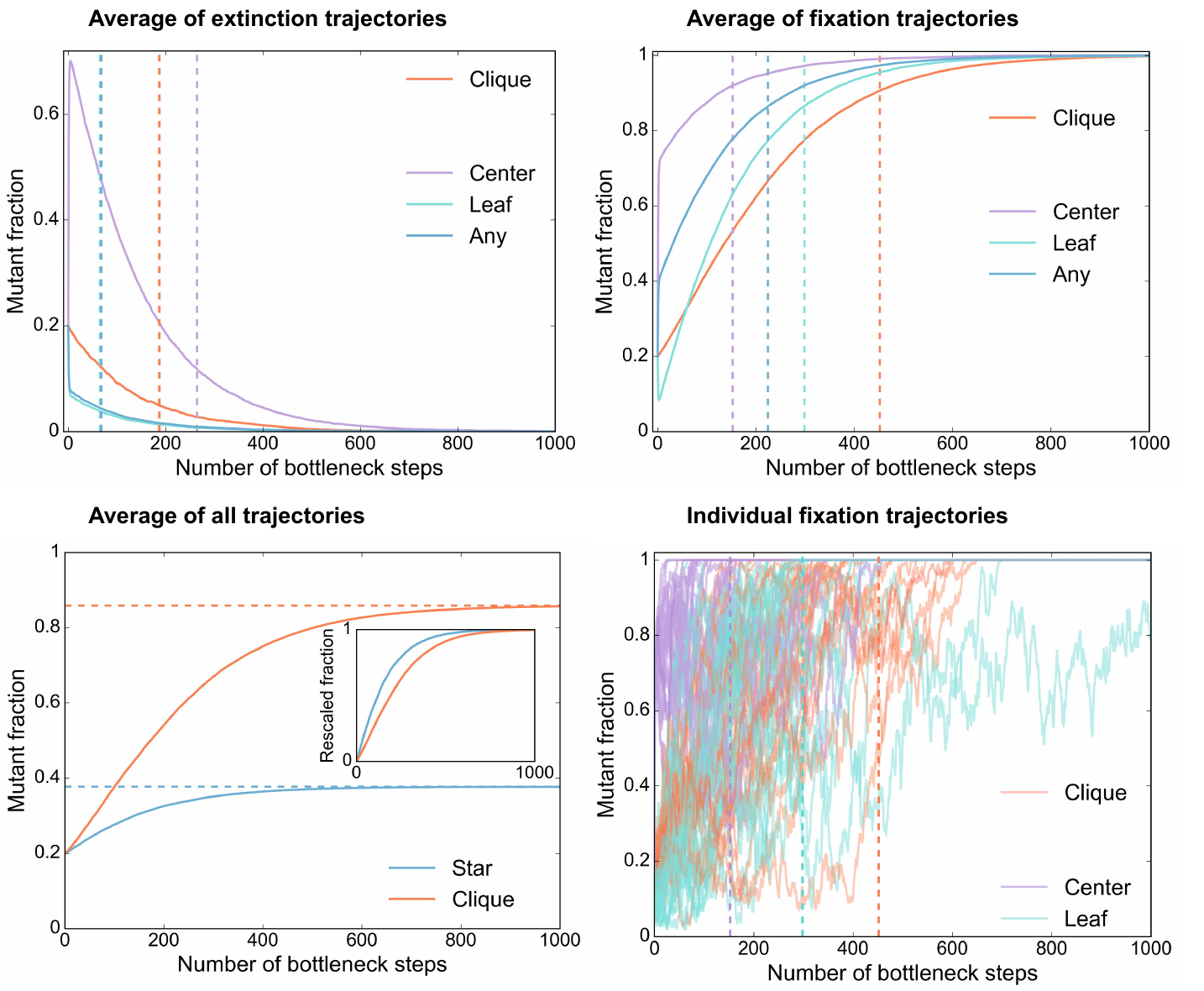}

     \caption{\textbf{Time evolution of the mutant fraction in the star.} Average of extinction trajectories (top left), of fixation trajectories (top right), of all trajectories (bottom left), and examples of single fixation trajectories (bottom right, 25 single trajectories are shown in each case). Vertical dashed lines indicate the average extinction times (top left panel) or fixation times (top right panel). In the inset of the bottom left panel, fractions are rescaled to begin at 0 and converge to 1, by employing the initial fraction and the fixation probabilities. As in Fig.~\ref{violin}, the star and clique comprise $D=5$ demes of size $B=100$ each, and  are initialized with one fully mutant deme. For the star, this deme is either the center, or a leaf, or a uniformly chosen one (``Any").  The effective fitness advantage of the mutants is set to $st=0.01$. For the star, $m_I=0.05$ and $\alpha=1/10$. For the star, averages are calculated over 5000 trajectories in each case, except for the relatively rare event of fixation starting from a fully mutant center, where 1800 trajectories are used.  For the clique, we average over 5000 fixation trajectories and 1000 extinction trajectories.  Recall that fixation probabilities are the following: for the star, $\rho=0.38$ starting from a uniformly chosen mutant deme ($\rho=0.96$ [resp. $\rho=0.23$] starting from a mutant center [resp. leaf]); for the clique  $\rho=0.86$. }
    \label{traj}
\end{figure}

\paragraph{Time evolution of the mutant fraction.} In addition to the probability of fixation and to the extinction and fixation times, evolution experiments in spatially structured populations also give access to the time evolution of mutant fraction before extinction or fixation happens. In Fig.~\ref{traj}, we show the theoretical prediction for these trajectories. We first show the average of extinction and fixation trajectories (top panels), and the average of all trajectories (bottom left panel). We observe that for the star starting from a fully mutant deme, the first bottleneck step yields a large variation of mutant fraction in both fixation and extinction trajectories, and that the average evolution is then smoother. The abrupt change at the first bottleneck is associated to the strong migration asymmetry ($\alpha=1/10$) in the star. Accordingly, it is not observed in the clique. The comparison of extinction (resp.\ fixation) trajectories in Fig.~\ref{traj} is in line with the comparison of average extinction (resp.\ fixation) times. For instance, fixation is accelerated starting from a fully mutant deme in the star compared to the clique. However, comparing overall average trajectories (including those leading to extinction and to fixation, see Fig.~\ref{traj}, bottom left panel) is more complex, since they converge toward intermediate mutant fractions corresponding to fixation probabilities. Thus, the acceleration effect is not easy to see from such average trajectories, unless the fractions are rescaled as in the inset.
In the regime considered here, fluctuations are important, as seen in individual fixation trajectories (see Fig.~\ref{traj}, bottom right panel). This entails that a sufficient number of replicates is needed to observe the acceleration effect (see our discussion above regarding extinction and fixation times).

\paragraph{Further mutations.} Since these experiments focus on the spread of a preexisting mutant strain, we need to ensure that there are few \textit{de novo} mutants. Here, with $B=100$, if $t=\log(100)$ as in~\cite{Chakraborty23} (which, for $st=0.01$, yields $s=2\times 10^{-3}$), the number of cell division events per growth phase is about $10^4$ in each deme. With a total mutation rate of $10^{-3}$ per genome and per bacterial replication~\cite{Lee12,Sung12,Robert18}, this yields about 10 \textit{de novo} mutation events per growth phase in each deme, which is not negligible. Spontaneous mutations generally have small fitness effects and are weakly deleterious to neutral~\cite{Robert18}, meaning that most of them they will be lost due to selection and drift. Nevertheless, to decrease the number of \textit{de novo} mutations during the experiment, an interesting possibility is to choose shorter growth times $t$. Keeping $st=0.01$, but taking $t=\log(10)$ instead of $t=\log(100)$ (and taking $s=4\times 10^{-3}$) reduces the number of \textit{de novo} mutation events to about 1 per growth phase in each deme. Note that reducing $t$ also reduces the total experiment duration: $t=\log(10)$ should correspond to performing about 10 serial passages per day instead of one, meaning that 1000 bottlenecks (the longest useful duration here) can be realized in 100 days. While performing frequent serial passages can be a challenge with manual methods, it could be achieved e.g.\ using a robot~\cite{Kryazhimskiy12}. As a whole, if maintained for 1000 bottlenecks, our proposed experiment comprises $5\times 10^6$ division events with $t=\log(10)$ and $5\times 10^7$ division events with $t=\log(100)$. This is much smaller than in each experiment performed in~\cite{Chakraborty23}, where $B=10^7$ and $t=\log(100)$ yields about $10^9$ divisions per deme and per growth phase, which gives more than $10^{10}$ division events total during an experiment.
If the selective advantage of the mutant of focus is ensured by antibiotic resistance as in~\cite{Chakraborty23}, spontaneous mutations conferring resistance may perturb the proposed experiments. Assuming that 100 possible single mutations increase resistance~\cite{Card21} yields a total mutation rate to resistance on the order of $10^{-8}$ per bacterial division, meaning that resistance mutations should remain rare in our proposed experiments.

\section*{Discussion}

Here, we used our recent model of spatially structured populations on graphs with serial passages~\cite{Abbara23} to quantitatively analyze recent evolution experiments and to propose future ones. 
Note that our model aims to reproduce the experimental protocol of~\cite{Chakraborty23} more closely than the model presented there. The latter is an agent-based model that considers much smaller demes than in actual experiments, and starts with only one mutant while experiments are started with substantial numbers of mutants. The main results of the model in~\cite{Chakraborty23} are obtained without explicit modeling of serial passage, although this aspect is accounted for in a variant of the model.
Conversely, our model considers demes with the same bottleneck size as in the experiments, and describes the serial passage protocol used in them. Indeed, we implement local growth of bacteria in each subpopulation, followed by a migration and dilution event. 

We predict no amplification or suppression of selection with the parameters used in~\cite{Chakraborty23}. We suggest using smaller bottleneck sizes and effective mutant fitness advantages, and starting from fully mutant demes, in the design of future experiments investigating possible amplification or suppression of natural selection by spatial structure. In this regime, we predict that the star with asymmetric migrations suppresses natural selection and can accelerate mutant fixation and extinction.

Our predictions in the regime we deem most promising for future experiments were obtained from numerical simulations. The key points, namely suppression of selection and acceleration of fixation and extinction for the star with small $\alpha$, are consistent with our analytical predictions in the branching process regime~\cite{Abbara23}. Note however that the conditions considered here are not in the branching process regime, most prominently because we start from one fully mutant deme. This shows the robustness of these trends. Recall that in evolutionary graph theory, the star can amplify or suppress natural selection depending on the update rule ~\cite{lieberman2005evolutionary,Kaveh15,Hindersin15,Pattni15}. We previously showed that with a deme on each node, the star amplifies selection for rare migrations and small fitness advantages, if incoming migrations to the center are stronger than outgoing ones~\cite{marrec2021}, but even then, it becomes a suppressor of selection for more frequent migrations~\cite{Abbara23}. However, the rare migration regime is problematic for experiments because it requires reproducibly handling very small numbers of cells (most migration-dilution events would involve zero exchange in this regime). In addition, it gives very slow dynamics, thus demanding extremely long experiments. This is why we advocate for a focus on evidencing suppression of selection. Note that we predict suppression when mutants are started in a uniformly chosen deme (which is the usual condition in which suppression or amplification are defined). However, we also predict that starting from a mutant center in the star with small $\alpha$ yields larger fixation probabilities for beneficial mutants than in the well-mixed population. Thus, experimental studies aiming to evidence increased fixation of beneficial mutants could focus on this specific case.

The experiments we proposed require precise serial passage with small bottlenecks on relatively long timescales. Another potential difficulty is the focus on mutants with relatively small advantages. However, we expect that they can be realistically achieved, e.g.\ using robots for serial passage~\cite{Kryazhimskiy12,Horinouchi14,Maeda20}. With the parameters we propose, the number of experimental replicates needed to see a significant effect remains modest, and \textit{de novo} mutants are not a strong concern. Beyond the specific experiments we suggested, our model can be simulated to predict experimental results for mutant fate in spatially structured populations on graphs undergoing serial passage. It thus connects models to experiments in a quantitative way.

Our model could be extended to include additional possible ingredients of experiments. For antibiotic resistant mutants, it could be relevant to model the mode of action, e.g.\ to distinguish bacteriostatic and bactericial actions. While birth and death can be lumped into an effective growth rate in exponential growth, this might matter with more complex growth models. Besides, specific experimental protocols can be effectively close to hard selection or to soft selection. This depends e.g.\ on whether a fixed dilution factor or a fixed bottleneck size is set~\cite{Mahrt21}, on whether stationary state is reached or not, and if it is, on whether carrying capacities are the same for different strains or not. Complete growth curves for both the mutant and the wild-type in the experimental conditions would provide valuable information about the latter points. Here, considering soft or hard selection  does not modify key conclusions, as in Refs.~\cite{Rouhani93,Barton93}. However, this point can be impactful e.g.\ for cooperative mutants exhibiting frequency-dependent selection~\cite{chuang2009simpson,Melbinger10,Cremer12,Cremer19}. 

From a theoretical point of view, our numerical simulations show that the main conclusions obtained in the branching process regime~\cite{Abbara23}, notably the pervasive suppression of selection associated to frequent asymmetric migrations, hold beyond it. This calls for further analytical study of our model of spatially structured populations. In particular, it will be interesting to develop a diffusion approximation~\cite{kimura_diff64,Ewens79,aurell19} for our model of spatially structured populations, as was done for structured Wright-Fisher models~\cite{Lessard07,Burden18}. This would also open the way to develop exact simulation algorithms~\cite{jenkins17, garcia-pareja21}.  

Another interesting theoretical point is that our model with demes on graph nodes allows acceleration of fixation with respect to the clique. In contrast, the clique has the shortest fixation time in evolutionary graph theory models with one individual per node~\cite{hindersin2013fixation,hindersin2014counterintuitive,Askari2015}. Interestingly, it was recently shown that in these models, higher-order motifs allow to tune fixation times~\cite{kuo2024evolutionary}. Thus, an investigation of fixation times in different graphs in our model would be of interest.

Here, our focus has been on the fate of mutants that are introduced at the beginning of experiments, i.e.\ of preexisting mutants. It would be interesting both theoretically and experimentally to consider longer evolutionary trajectories~\cite{Sharma22}, and to include \textit{de novo} mutants, which may appear at any time during the growth phase~\cite{Wahl01,Wahl02,LeClair18,Lin20,Freitas21,Gamblin23}. Further extensions include addressing environmental gradients~\cite{Zhang11,Greulich12,Hermsen12,Baym16}, changing environments~\cite{Mustonen08,Ashcroft14,Hufton16,Hufton19,Marrec20a,Marrec20b,Marrec23}, possible deme extinctions~\cite{Barton93}, cooperative interactions~\cite{ohtsuki2006evolutionary,Gokhale18,Moawad24}, and connecting to models of expanding populations~\cite{Hallatschek07,Hallatschek08}. 

\section*{Model and methods}
\label{Model_methods}
\subsection*{Model} 

We model a spatially structured population as a connected graph with $D$ nodes, each one comprising a well-mixed deme. Our model was first presented in~\cite{Abbara23}, but we explain it here for the sake of completeness, and we explicitly make the link with the experiments performed in~\cite{Chakraborty23}. Migration probabilities $m_{ij}$ are defined between any pair of demes $(i,j) \in \{1,...,D\}^2$, including the case $i=j$ which describes individuals that stay in the same deme. We consider two types of individuals: wild-types and mutants. They differ by their exponential growth rates, which are respectively $f_W=1$ for wild-types (taken as reference) and $f_M=1+s$ for mutants. Thus, $s$ is the relative fitness advantage of mutants compared to wild-types. The dynamics of this spatially structured population, modeling serial passage with migrations, comprises successive growth and sampling phases. Migrations are implemented during the sampling phases. A different version of this model that implements hard selection is detailed in the Supplement. In the batch culture experiments in structured populations of~\cite{Chakraborty23}, each deme corresponds to a well that can contain wild-type and mutant bacteria. The latter are resistant to the antibiotic ciprofloxacin, which is present in the culture medium. This is the origin of the fitness advantage of mutants compared to wild-types.

As depicted in the top panel of Fig.~\ref{model_structures}, starting from a bottleneck, an elementary step of the dynamics includes two successive phases detailed below, that result in a new bottleneck. We perform stochastic simulations of this model. 

\paragraph{Growth phase.} First, a phase of local exponential growth happens in each deme during a fixed amount of time $t$. Let us denote by $W_i$ the number of wild-types and  by $M_i$ the number of mutants initially present in deme $i$.
At the end of the growth phase, the numbers become $W_i' = W_i \exp (f_W t)= W_i \exp (t)$ and $M_i'=M_i \exp (f_M t)=W_i \exp[(1+s) t]$, and the total size of the population in deme $i$ is $N_i'=M_i' + W_i'$. During the growth phase, the fraction of mutants changes from $x_i = M_i/N_i$, with $N_i=M_i+W_i$, to $x_i'=M_i'/N_i'= x_i e^{st} / [1+x_i(e^{st}-1)]$. Importantly, the impact of the selective advantage of mutants is encoded by the product $st$ of relative selective advantage $s$ and growth time $t$. We thus call $st$ the effective fitness advantage.

This phase corresponds to starting with fresh culture medium containing bacteria, and letting them grow. 

\paragraph{Dilution and migration phase.} Then, a phase that simultaneously incorporates dilution and migration is carried out by performing independent samplings. The number of mutants (respectively wild-types) sent from a deme $i$ to a deme $j$ is sampled from a binomial distribution with $N_i'$ trials, and a probability of success $m_{ij} x_i' B/N_i'$ (respectively $m_{ij} (1-x_i') B/N_i'$). Each deme $j$ will thus receive an average number of $B \sum_i m_{ij}$ individuals, which gives its new bottleneck state. For simplicity, we generally assume that for all $j$, $\sum_i m_{ij}=1$, meaning that each deme has the same average bottleneck size $B$. 
This process amounts to drawing approximately $B m_{ij}$ bacteria from the total $N_i'$ bacteria present in deme $i$, and using them to form the new bottleneck state of deme $j$. 
Note however that for consistency with~\cite{Chakraborty23}, we performed the simulations and the calculations of Figs.~\ref{fixation_det} and~\ref{alpha3_small} assuming instead that $\sum_i m_{ji}=1$, meaning that each deme contributes to the next bottleneck by the same average amount $B$ (see~\cite{Abbara23} for details).

Experimentally, in~\cite{Chakraborty23}, structured populations were transferred and placed in fresh culture medium right after dispersal among subpopulations.

\subsection*{Parameter choices matching the experiments performed in~\cite{Chakraborty23}}

In~\cite{Chakraborty23}, structured populations of $D=4$ demes, each growing in a well of a multi-well plate, are considered. Their bottleneck sizes are $N_i\approx B \approx10^7$ colony-forming units in each well $i$. Cultures undergo daily transfer, and grow to a final size of $N'_i\approx 10^9$ colony-forming units. The duration $t$ of exponential growth that should be taken in our model to match these experiments satisfies $N'_i/N_i=e^t=100$ (considering wild-types), yielding $t=\log(100)$. 

The competitive advantage of mutants in the experimental conditions of~\cite{Chakraborty23} is reported to be about 20\%. This competitive advantage was measured via head-to-head competition starting with equal abundances of mutants and wild-types. This corresponds to measuring \cite{Wiser15}
\begin{equation}
\frac{\log(M'_i/M_i)}{\log(W'_i/W_i)}=\frac{f_M}{f_W}=1+s\,.
    \label{eq:fitness}
\end{equation}
Thus, $s=0.2$ in these experiments.

Migration probabilities in~\cite{Chakraborty23} range from 0.001\% to 30\%, i.e. from $10^{-5}$ to 0.3. For the star, different asymmetries are considered, namely $\alpha=1/3$ (referred to as ``OUT$>$IN" or ``amplifier'' in~\cite{Chakraborty23}),  $\alpha=1$ (``balanced amplifier" in~\cite{Chakraborty23}) and $\alpha=3$ (``IN$>$OUT" in~\cite{Chakraborty23}). The clique is also considered, and referred to as ``well-mixed'' in~\cite{Chakraborty23}. Note that for completeness, we also consider the well-mixed population with no deme structure and the same total bottleneck population size as the star and the clique. This case was not realized in~\cite{Chakraborty23}. In the case of the star, the exact way incoming migrations to the center were implemented was by constructing a mix from all three leaves, diluting it and putting it in the center together with some diluted bacteria from the center and with fresh culture medium and antibiotic. Outgoing migrations were implemented by diluting bacteria from the center and putting them in each leaf together with some diluted bacteria from that leaf and with fresh culture medium and antibiotic. We further note that with the exact migration schemes used in~\cite{Chakraborty23}, the star with $\alpha=1$ does not strictly satisfy $\forall j,\,\,\sum_im_{ij}=1$ or $\forall j,\,\,\sum_im_{ji}=1$ (but the clique satisfies $\forall j,\,\,\sum_im_{ij}=\sum_im_{ji}=1$). Meanwhile, the star with $\alpha=1/3$ satisfies $\forall j,\,\,\sum_im_{ij}=1$, and the star with $\alpha=3$ satisfies $\forall j,\,\,\sum_im_{ji}=1$. However, we showed in~\cite{Abbara23} that choosing one or the other of these conventions does not qualitatively change how the star impacts mutant fixation. We checked that it is also the case here, in our reproductions of the setup of~\cite{Chakraborty23}. 

To assess the robustness of our approach, we also introduce a different version of our model that implements hard selection, and where mutant demes reach saturation in the growth phase. In this variant of the model, parameters can also be chosen to best match the experiments (see Section~\ref{Section_hard_selection} in the Supplement). 

Even with our efforts to match the experimental conditions of~\cite{Chakraborty23}, there remain some differences. In particular, the experimental results show a quicker increase of mutant fraction in the population. This might be due to the mutants having a higher fitness advantage than 0.2, and to the fact that the antibiotic induces cell death. 
Besides, there are stronger fluctuations (including some decreases of mutant fraction) in the experimental results than in our stochastic simulations. This might be due e.g.\ to antibiotic concentration variability in the experiments, or to the difficulty of diluting subpopulations in a perfectly reproducible manner.
An experimental study of these elements would be very interesting.

\subsection*{Fixation probability in a well-mixed population undergoing serial passage}

In our growth and dilution model, the probability of fixation starting from an initial number $n$ of mutants at a bottleneck in a well-mixed population with bottleneck size $B$ reads~\cite{Abbara23}
\begin{equation}
    \rho=\frac{1-e^{-2n st}}{1-e^{-2B st}}\,.
    \label{fix}
\end{equation}
This result is obtained in the diffusion approximation, assuming $B\gg 1$ and $|s|t\ll 1$. Our serial dilution model in general, and this formula in particular, exhibit a strong formal similarity with the Wright-Fisher model~\cite{Ewens79}. The effective fitness advantage $st$ in our model plays the part of the fitness advantage in the Wright-Fisher model. Note that with the parameter values matching~\cite{Chakraborty23}, we are not in the regime where $|s|t\ll 1$. However, the diffusion approximation is often reasonable even beyond this regime~\cite{Brautingam22}.

\subsection*{Deterministic model}

In addition to our complete stochastic model described above, we propose a deterministic model based on a recurrence relation. This model will be valid when fluctuations in the mutant fractions in each deme can be neglected. This holds in the regime of large populations, large initial numbers of mutants and large migration probabilities. 
At a given bottleneck $n$, let us denote by $x_{i,n}$ the mutant fraction in each deme $i$ with $1\leq i\leq D$. Neglecting fluctuations, the mutant fraction $x_{i,n+1}$ in deme $i$ at the next bottleneck can be expressed as:
\begin{equation}
    x_{i,n+1}=\dfrac{\sum_{j=1}^D m_{ji} x'_{j,n}}{\sum_{j=1}^D m_{ji}},
    \label{recurrence}
\end{equation}
where $x'_{j,n}= x_{j,n} e^{st} / [1+x_{j,n}(e^{st}-1)]$ denotes the mutant fraction in deme $j$ after the growth phase. The mutant fraction in each deme at each bottleneck can be obtained by iterating Eq.~\ref{recurrence}, starting from the appropriate initial conditions.

\subsection*{Number of experimental replicates needed for the effect of spatial structure to be significant}

\paragraph{Fixation probabilities.} We compute the Wilson confidence interval~\cite{Wilson27} for the fixation probability $\rho$, estimated as the proportion $\hat{\rho}$ of experiments where mutants fix in the population. The bounds $\rho_\pm$ of the Wilson confidence interval on the estimated binomial proportion $\hat{\rho}$ are
\begin{equation}
\rho_\pm=\frac{1}{1+\frac{z^2}{n}}\left(\hat{\rho}+\frac{z^2}{2n}\pm z\sqrt{\frac{\hat{\rho}(1-\hat{\rho})}{n}+\frac{z^2}{4n^2}}\right),
\label{Wilson}
\end{equation}
where $z$ is the appropriate quantile from the standard normal distribution and $n$ is the number of experimental replicates performed. Using Eq.~\ref{Wilson}, and using our precise estimates of $\rho$ from numerical simulations instead of the experimental estimates $\hat{\rho}$, we find that for the parameters considered, $n=16$ replicates are sufficient to detect a significant difference between the fixation probability in the star starting from a uniformly chosen mutant deme and the one in the clique. Here, significant difference is understood in the stringent sense of non-overlapping confidence intervals, at a $95\%$ confidence level (in this case, $z=1.96$ for a two-sided interval). Note that, compared to the Wald confidence interval for binomial proportions, the Wilson confidence interval has a better average coverage~\cite{Brown01}, in particular when $\rho$ is somewhat close to 0 or 1. 

\paragraph{Fixation times.} The bounds $\tau_\pm$ of the confidence intervals for the experimentally estimated average fixation times $\hat{\tau}$ are computed as:
\begin{equation}
\tau_\pm=\hat{\tau} \pm t_{(n-1)} \frac{\hat{\sigma}}{\sqrt{n}}\,,
\label{CI}
\end{equation}
where $\hat{\sigma}$ is the experimental estimate of the standard error of the fixation time, and $t_{(n-1)}$ is the appropriate quantile from a $t$-distribution with $n-1$ degrees of freedom. Here, we use the average values and standard errors of fixation times obtained from simulations and corresponding to the distributions depicted in Fig.~\ref{violin} instead of experimental estimates: for the star, starting from the center, we use $\hat{\tau}=150.67$ and $\hat{\sigma}=125.48$, and for the clique, we use $\hat{\tau}=451.24$ and $\hat{\sigma}=198.25$. Using these values, we find that for the parameters considered, $n=7$ experimental replicates are sufficient to detect a significant difference between the average fixation time in the star starting from a mutant center and the one in the clique. Here, significant difference is understood as non-overlapping confidence intervals, at a $95\%$ confidence level ($t_{(6)}=2.45$ for a two-sided interval). Note that the same calculation can be used for extinction times.

\section*{Code availability statement}
Python code for our numerical simulations is freely available at \url{https://github.com/Bitbol-Lab/Structured_pop}.

\section*{Acknowledgments}
The authors thank Partha Chakraborty and Rees Kassen for private communications about the details of their experiments, as well as Ana-Hermina Ghenu and two anonymous reviewers for comments that allowed to substantially improve this paper.
 A.-F.~B. thanks Loïc Marrec for helpful discussions. This project has received funding from the European Research Council (ERC) under the European Union’s Horizon 2020 research and innovation programme (grant agreement No.~851173, to A.-F.~B.).

\newpage
\appendix

\begin{center}
\LARGE{\textbf{Supplementary material}}
\end{center}
\vspace{0.2cm}

\renewcommand{\thesection}{S\arabic{section}}
\renewcommand{\thefigure}{S\arabic{figure}}
\setcounter{figure}{0}
\renewcommand{\thetable}{S\arabic{table}}
\setcounter{table}{0}
\renewcommand{\theequation}{S\arabic{equation}}
\setcounter{equation}{0}

\section{Hard selection variant of the model}
\label{Section_hard_selection}

In the main text, we focused on a serial dilution model that implements soft selection (see ``Model and methods''). Indeed, the total number of migrants sent out from a deme at the dilution step does not depend on its size after growth. In addition, the bottleneck size of each deme has a fixed average value $B$. 

Here, we propose a variant of the model that relies on hard selection. As usual, migration probabilities $m_{ij}$ are defined between any pair of demes, wild-types have fitness $f_W=1$ and mutants have fitness $f_M = 1+s$. While all demes are initially of size $B$, they now also have a carrying capacity $K_{sat}$ that cannot be exceeded during the growth phase. We use a fixed dilution factor $d$. An elementary step of the model still includes a growth and dilution phase, detailed below.

\paragraph{Logistic growth phase.} A phase of logistic growth happens in each deme for a fixed time $t$. We denote by $M_i(0)$ the number of mutants and by $W_i(0)$ the number of wild-types initially present in deme $i$. For $\tau \in [0, t]$, these numbers grow following:
\begin{align}
    \dfrac{d W_i(\tau)}{d\tau} &= W_i(\tau) f_W \left( 1 - \dfrac{W_i(\tau) + M_i(\tau)}{K_{sat}}\right),\\
    \dfrac{d M_i(\tau)}{d\tau} &= M_i(\tau) f_M \left( 1 - \dfrac{W_i(\tau) + M_i(\tau)}{K_{sat}}\right).
\end{align}
At the end of the growth phase, deme $i$ contains $M_i'$ mutants and $W_i'$ wild-types, and its size $N_i'=M_i' + W_i'$ cannot exceed $K_{sat}$. The fraction of mutants in deme $i$ at the end of the growth phase is $x_i'= M_i'/N_i'$.

\paragraph{Dilution and migration phase.} The dilution phase is similar to the one detailed in ``Models and methods", but each deme is now diluted with a fixed factor $d$. Specifically, the number of mutants (respectively wild-types) that migrate from deme $i$ to deme $j$ is sampled from a binomial distribution with $N_i'$ trials and probability of success $x_i' m_{ij} / d$ (respectively $(1-x_i')m_{ij}/d$). The mean number of mutants (resp.\ wild-types) sampled in this way is proportional to $M'_i$ (resp.\ $W'_i$). Thus, demes that reach a larger size at the end of the growth phase send out proportionally more individuals. This implements hard selection. 

\paragraph{Choice of parameters.} In the experiments of Ref.~\cite{Chakraborty23}, at the end of the growth phase, there was a 30 to 50\% reduction in optical density in fully wild-type demes with respect to fully mutant demes~\cite{ChakrabortyP}. Besides, dilution was performed with a fixed dilution factor $d=100$~\cite{Chakraborty23}. The initial size of all demes at the beginning of the experiment is $B=10^7$. For a given value of mutant fitness $s$, we choose the carrying capacity $K_{sat} $ such that:
\begin{itemize}
    \item Starting with a bottleneck size $B=10^7$, the size of a fully wild-type deme increases by a factor $100$ during the growth phase. Therefore, in the absence of migrations, deme size would remain equal to $B$ at the next bottleneck.
    \item A fully mutant deme, initialized at bottleneck size $B=10^7$, increases in size through successive bottlenecks. After a few steps of the dynamics, it reaches carrying capacity $K_{sat}$ at the end of each growth phase.
    \item The size reached by a fully wild-type deme at the end of the growth phase is $40\%$ smaller than the carrying capacity, i.e. $100 B = 0.6 K_{sat}$.
\end{itemize}
For $s=0.2$, the corresponding carrying capacity is $K_{sat}\approx 1.9 \times 10^9$ and the growth time is $t \approx 5.3$. For $s=0.3$, we obtain $K_{sat}\approx1.8 \times 10^9$ and $t \approx 5.4$.
With these parameter choices, we reproduce to the best of our knowledge the conditions of the experiments conducted in~\cite{Chakraborty23}. Note however that a more precise match could be obtained if complete growth curves for both the mutant and the wild-type were available in the experimental conditions. Another useful point would be to measure division rate and death rate independently of each other in the presence of antibiotic. This would allow more detailed modeling of antibiotic action.

\begin{figure}[htbp]
    \centering
\includegraphics[scale=0.37]{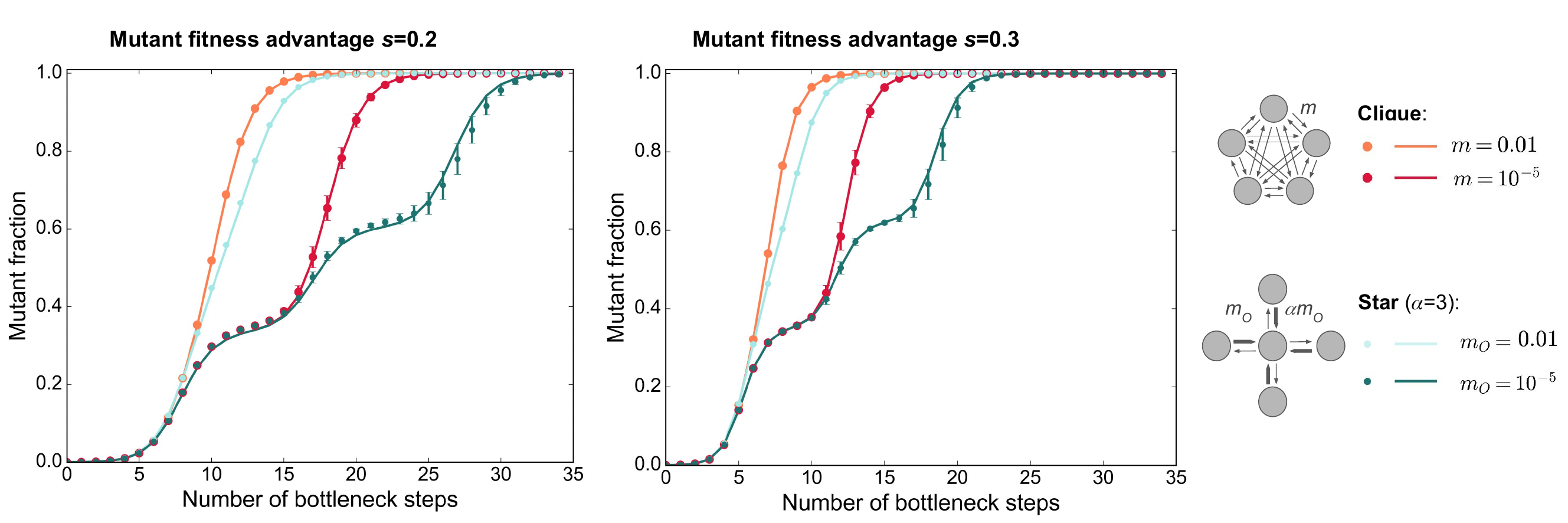}

\caption{\textbf{Growth of the mutant fraction in the star and the clique in the hard selection model.} As in~\cite{Chakraborty23}, we consider a clique and a star with $D=4$ demes of initial bottleneck size $B=10^7$, where $10^4$ mutants with fitness advantage $s=0.2$ (left) and $s=0.3$ (right) are initially placed in one leaf for the star structure, any leaf for the clique. The dilution factor is $d=100$. In each panel, the carrying capacity and growth time are chosen to obtain a 40\% optical density between fully wild-type demes and fully mutant demes after growth. We consider different migration probabilities $m$ for the clique and $m_I$ for the star, satisfying $m=m_O$ and $\alpha=m_I/m_O=3$ for the star. Markers are simulation results averaged over 100 trajectories, and error bars report the standard deviation. Note that all trajectories resulted in mutant fixation. Lines show the predictions from our deterministic model, adapted to the hard selection version of the model.
\label{hard_selection_fig}}
\end{figure}

In Fig.~\ref{hard_selection_fig}, we show the growth of mutant fraction in different structures, in a case matching the top right panel from Fig.~\ref{fixation_det} in the main text, but in the hard selection variant of the model. Fig.~\ref{hard_selection_fig} can be directly compared to Fig.~3 from Ref.~\cite{Chakraborty23}, in the ``IN$>$OUT'' setting. As these experiments, we focus on the case of mutants initially introduced in the leaf.
Migration rates of $m=0.01$ for the clique and $m_O=0.01$ for the star correspond to the ``High (1\%)'' migration intensities in~\cite{Chakraborty23}. Migration rates of $m=10^{-5}$ for the clique and $m_O=10^{-5}$ for the star correspond to the ``Low (0.001\%)'' case. We recover the same main conclusions in the hard selection variant of our model as those obtained with the soft selection model, presented in the main text. In both cases, the mutant fraction growth displays thresholds for small migration intensities, as demes are successively invaded by mutants. However, the mutant fractions reached at plateaus are different, because in our hard selection model, the bottleneck sizes of the demes depend on their composition. We still obtain a very good agreement between the deterministic model (adapted to include hard selection) and our simulation results. Again, we find that the mutant fraction does not grow faster in the star than in the clique.

Note that the mutant fraction grows faster in experiments from~\cite{Chakraborty23} than in our simulations. Taking a larger mutant fitness advantage in our model could account for this difference.
As shown in the right panel of Fig.~\ref{hard_selection_fig}, setting $s=0.3$ accelerates the mutant fraction growth with respect to the left hand panel where $s=0.2$ (which was chosen to match the measurements of~\cite{Chakraborty23}).

\newpage

\section{Supplementary figures}

\begin{figure}[h!]
    \centering
\includegraphics[scale=0.338]{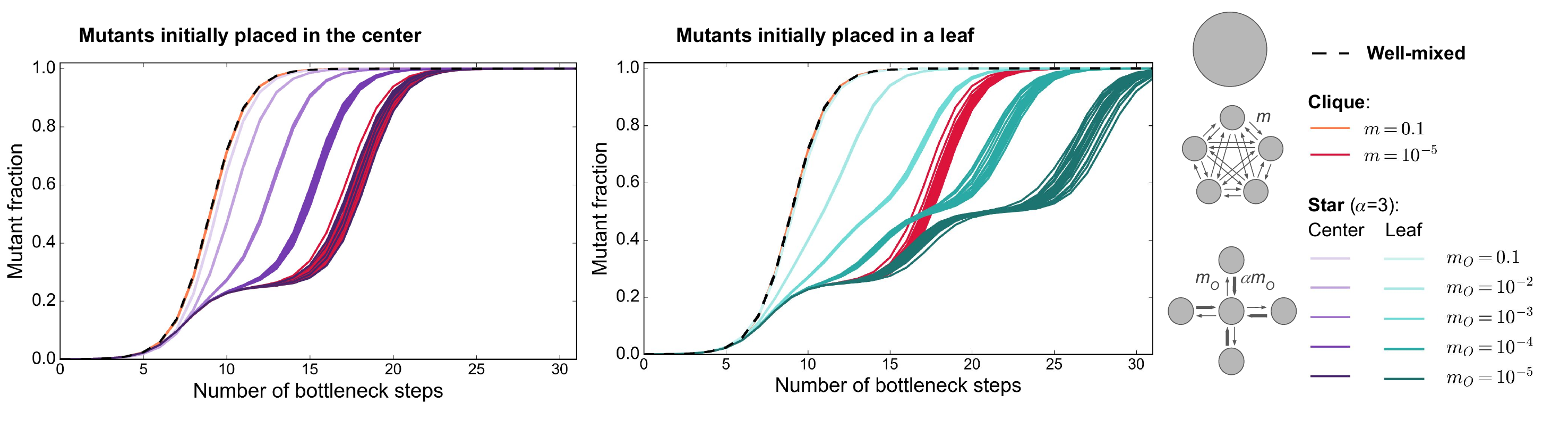}

\caption{\textbf{Growth of the mutant fraction in single trajectories in different structures.} As in Fig.~\ref{fixation_det} and in~\cite{Chakraborty23}, we consider a clique and a star with $D=4$ demes of bottleneck size $B=10^7$ each, where $10^4$ mutants with effective fitness advantage $st=0.2 \log(100)$ are initially placed in one deme. As a reference, we also consider a well-mixed population of total bottleneck size $DB=4 \times 10^7$, initialized with $10^4$ mutants. Left: mutants initially placed in the center of the star; right: mutants initially placed in a leaf of the star (the same trajectories are shown in both panels for the clique and the well-mixed population, and serve as references). In each case, we report the mutant fraction versus time (expressed in number of bottleneck steps). We consider the same migration probabilities as in Fig.~\ref{fixation_det}. Each line reports the result of a single stochastic simulation (i.e. a single trajectory), and 20 of them are shown in each case. Note that all of them result in mutant fixation.}
\label{fixation_det_leaf_center}
\end{figure}

\begin{figure}[htb!]
    \centering  
\includegraphics[scale=0.42]{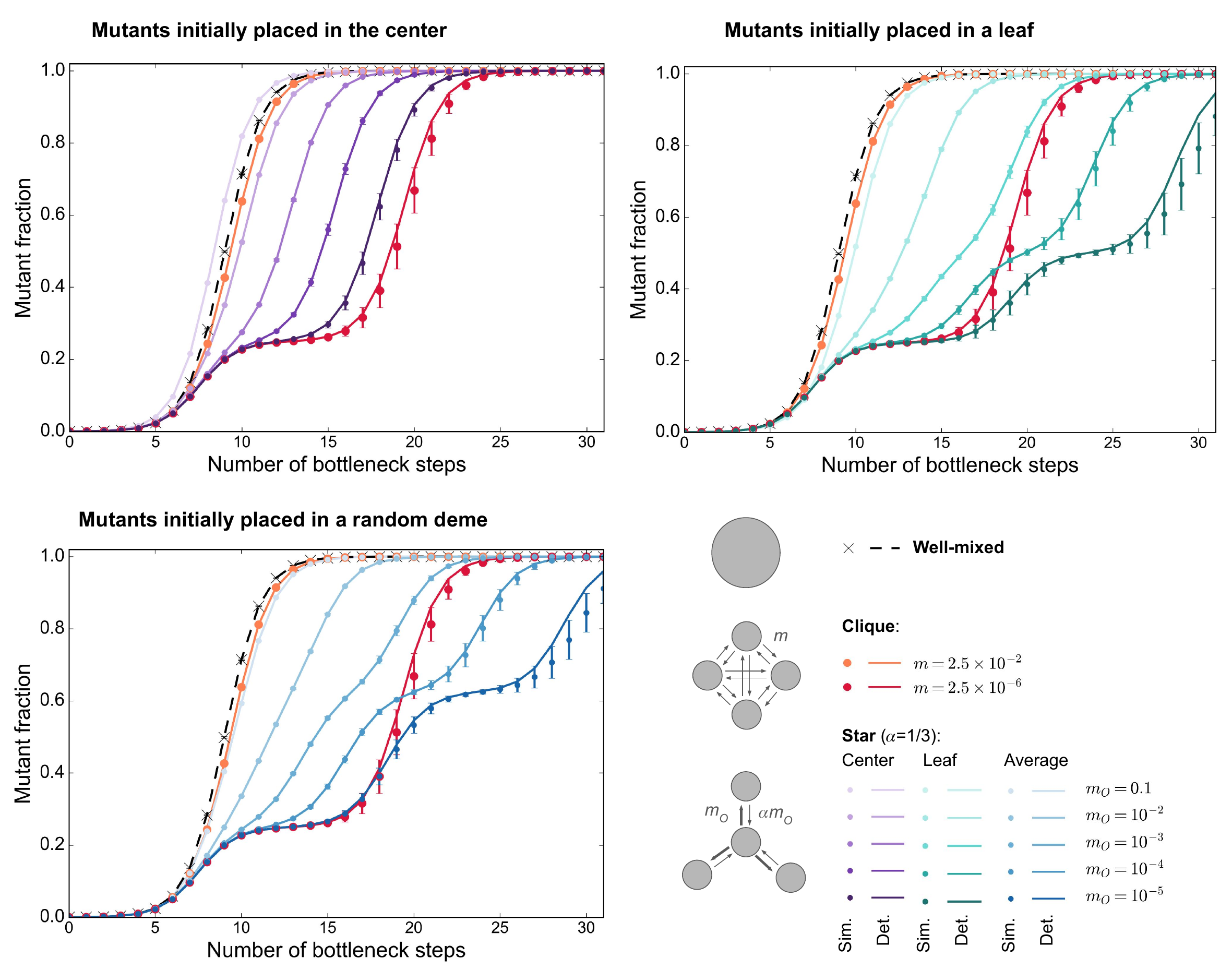}

    \caption{\textbf{Growth of the mutant fraction in different structures for $\alpha=1/3$.} As in~\cite{Chakraborty23} and in Fig.~\ref{fixation_det}, we consider a clique and a star with $D=4$ demes of bottleneck size $B=10^7$ each, where $10^4$ mutants with effective fitness advantage $st=0.2 \log(100)$ are initially placed in one deme. This deme is the center (top left), a leaf (top right), or chosen uniformly at random (bottom). As a reference, we also consider a well-mixed population of total bottleneck size $DB=4 \times 10^7$, initialized with $10^4$ mutants. In each case, we report the mutant fraction versus time (expressed in number of bottleneck steps). We consider different migration probabilities $m$ for the clique and $m_I$ for the star, keeping $\alpha=m_I/m_O=1/3$ for the star (recall that $\alpha=3$ in Fig.~\ref{fixation_det}). Markers are simulation results averaged over 100 trajectories, and error bars report the standard deviation. Note that all trajectories resulted in mutant fixation. Lines show the predictions from our deterministic model.}
    \label{alpha1/3}
\end{figure}

\begin{figure}[htb!]
    \centering
\includegraphics[scale=0.42]{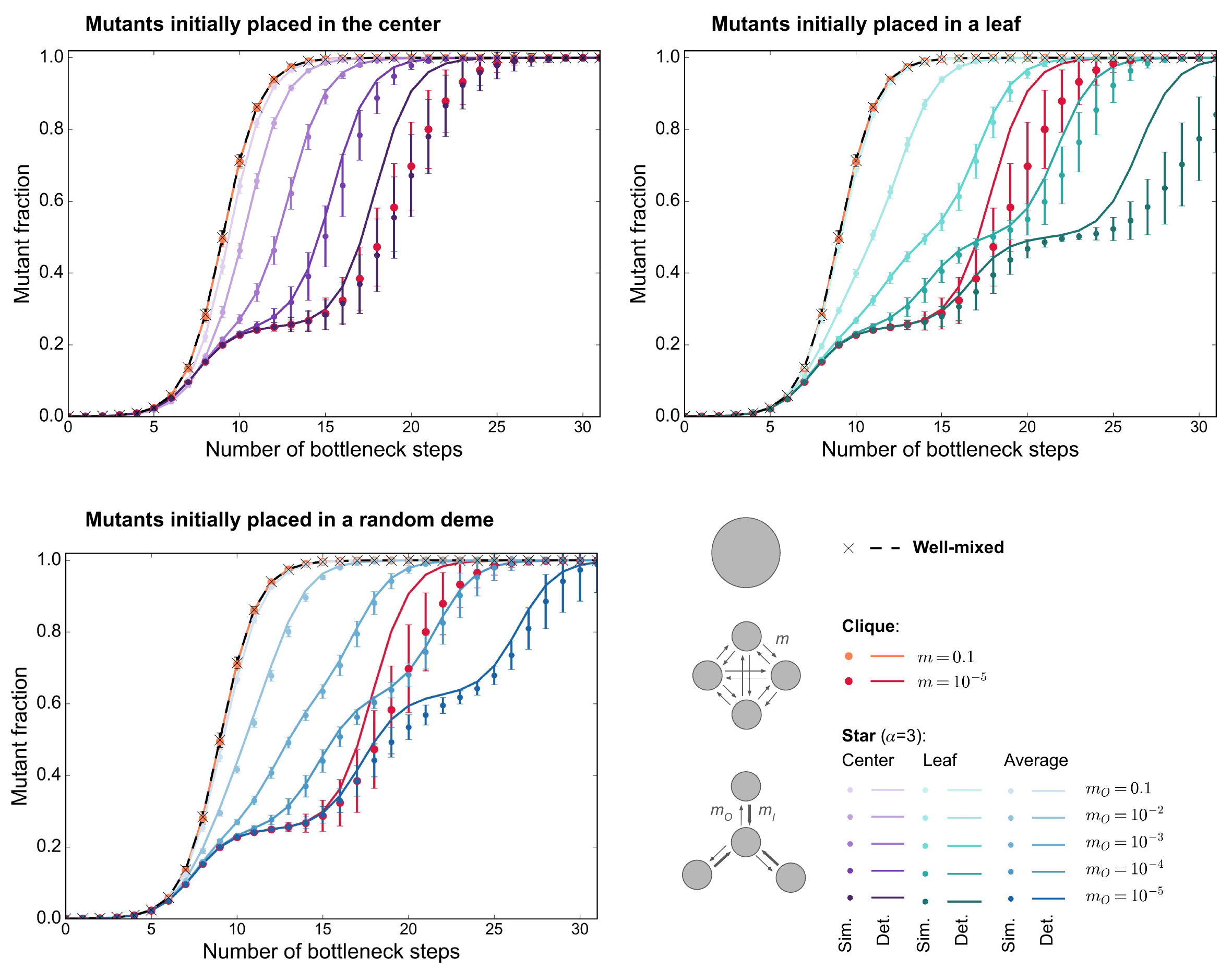}

    \caption{\textbf{Growth of the mutant fraction in different structures for $B=10^5$.} Same as Figs.~\ref{fixation_det} and~\ref{alpha1/3}, but for smaller bottleneck deme sizes $B=10^5$, and with $\alpha=3$.}
    \label{alpha3_small}
\end{figure}
\newpage

\begin{figure}[htbp]        
        \centering
\includegraphics[scale=0.41]{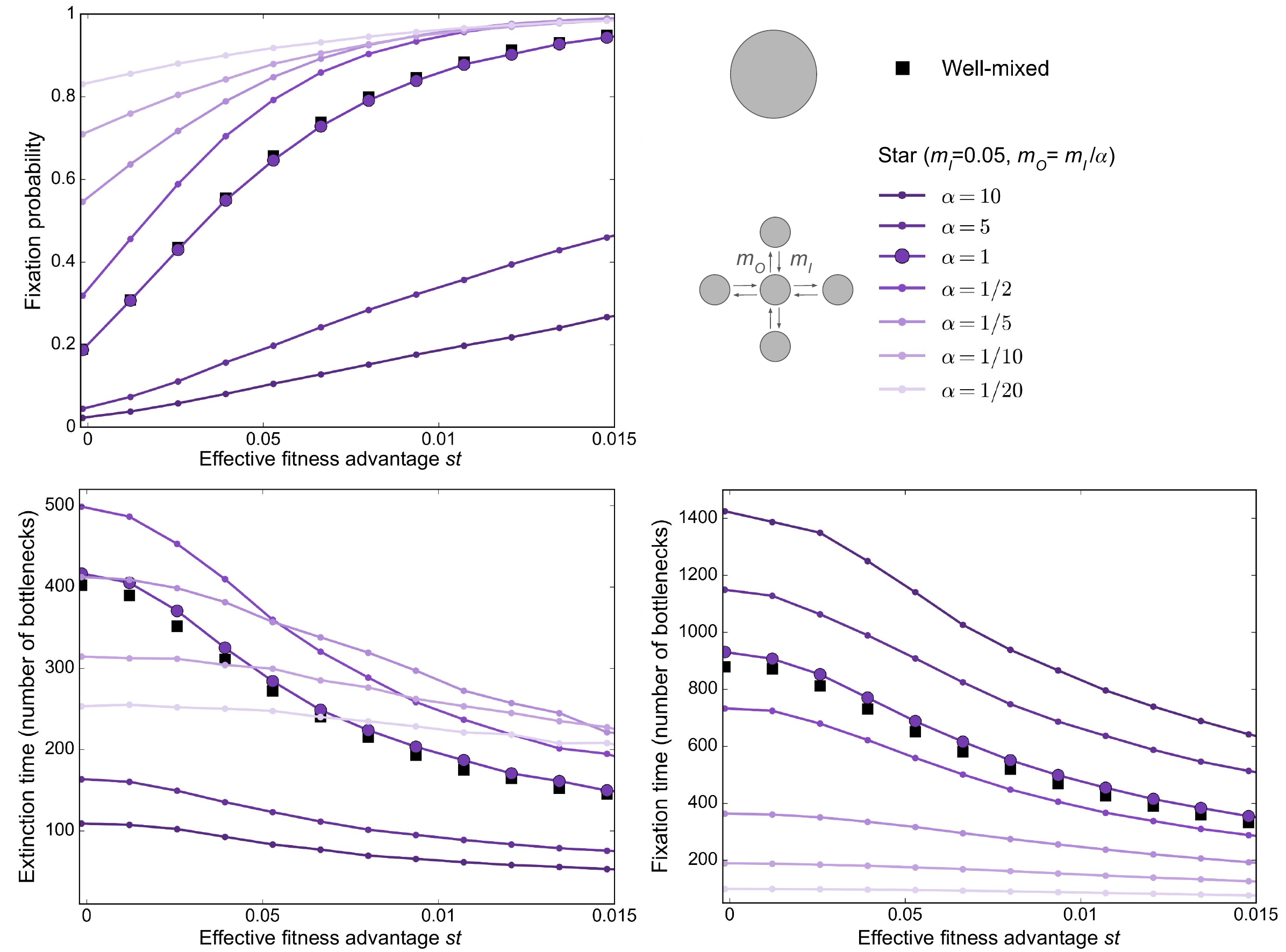}

    \caption{\textbf{Mutant fixation and extinction in the star, starting from a mutant center.} Mutant fixation probability (top), average extinction time (bottom left) and average fixation time (bottom right) are plotted as a function of the effective fitness advantage $st$ of the mutant. We consider a star with $D=5$ demes of size $B=100$, as in Fig.~\ref{fprobas_etimes_ftimes_full}, but it is initialized with a fully mutant center. For reference, we also consider a well-mixed population of size $DB=500$, initialized with 100 mutants. We take different values of $\alpha=m_I/m_O$, always with $m_I=0.05$, as in Fig.~\ref{fprobas_etimes_ftimes_full}. Markers are simulation results, obtained over at least 100,000 realizations. Lines linking markers are guides for the eye.}
    \label{fprobas_etimes_ftimes_fullcenter}
\end{figure}

\newpage

\begin{figure}[htbp]        
        \centering
\includegraphics[scale=0.41]{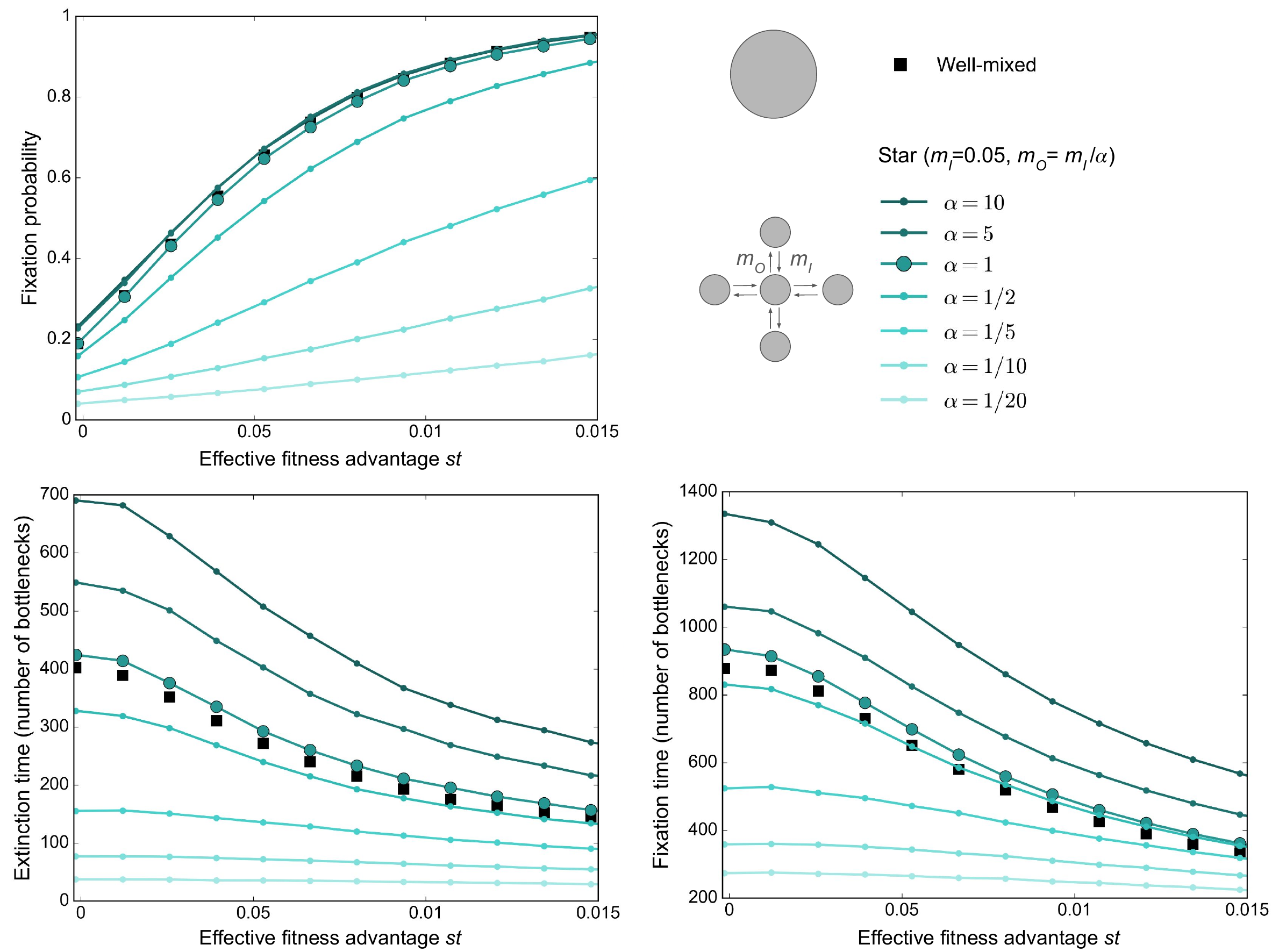}

        \caption{\textbf{Mutant fixation and extinction in the star, starting from a mutant leaf.} Same as Fig.~\ref{fprobas_etimes_ftimes_fullcenter}, but starting from a fully mutant leaf instead of a fully mutant center. Mutant fixation probability (top), average extinction time (bottom left) and average fixation time (bottom right) are plotted as a function of the effective fitness advantage $st$ of the mutant. We consider a star with $D=5$ demes of size $B=100$, as in Fig.~\ref{fprobas_etimes_ftimes_full}, but it is initialized with a fully mutant leaf. For reference, we also consider a well-mixed population of size $DB=500$, initialized with 100 mutants. We take different values of $\alpha=m_I/m_O$, always with $m_I=0.05$, as in Fig.~\ref{fprobas_etimes_ftimes_full}. Markers are simulation results, obtained over at least 100,000 realizations. Lines linking markers are guides for the eye.}
    \label{fprobas_etimes_ftimes_fullside}
\end{figure}

\newpage

\begin{figure}[htbp]        
        \centering
\includegraphics[scale=0.41]{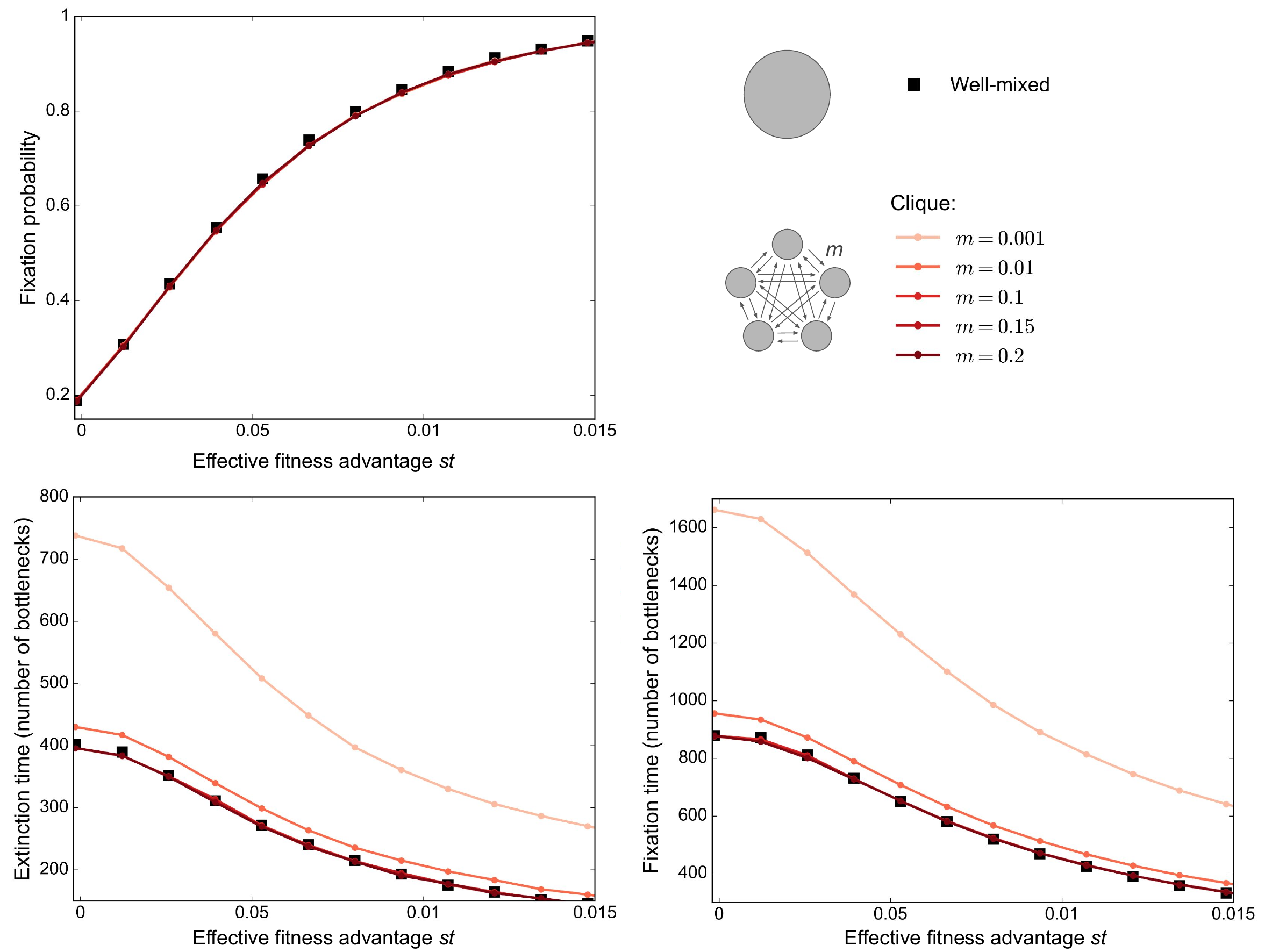}

        \caption{\textbf{Mutant fixation and extinction in the clique.} Mutant fixation probability (top), average extinction time (bottom left) and average fixation time (bottom right) are plotted as a function of the effective fitness advantage $st$ of the mutant. We consider a clique with $D=5$ demes of size $B=100$, initialized with a fully mutant deme, as in Fig.~\ref{fprobas_etimes_ftimes_full}. For reference, we also consider a well-mixed population of size $DB=500$, initialized with 100 mutants. We take different values of the migration probability $m$. Markers are simulation results, obtained over at least 100,000 realizations. Lines linking markers are guides for the eye.}
    \label{etimes_ftimes_full_clique}
\end{figure}

\clearpage
\newpage
\bibliographystyle{unsrt}
\bibliography{biblio_ALCAF}

\end{document}